\documentclass[12pt]{article}

\usepackage{graphicx}
\usepackage{multirow}
\usepackage{booktabs}
\usepackage{array}
\usepackage{booktabs,subcaption,amsfonts,dcolumn}
\usepackage{soul}
\usepackage{authblk} 
\usepackage{natbib} 
\usepackage{url} 
\usepackage{amsmath}
\newcommand{\blind}{0}
\usepackage{enumitem}
\usepackage[doublespacing]{setspace}
\usepackage{bm}
\usepackage{amssymb}
\usepackage[font=scriptsize]{caption} 
\usepackage[ruled,vlined]{algorithm2e}

\DeclareMathOperator{\E}{E}

\DeclareMathOperator*{\argmax}{argmax}

\newcolumntype{M}[1]{>{\centering\arraybackslash}m{#1}}

\usepackage{xcolor}

\usepackage{booktabs}

\makeatletter
\newcommand*{\addFileDependency}[1]{
  \typeout{(#1)}
  \@addtofilelist{#1}
  \IfFileExists{#1}{}{\typeout{No file #1.}}
}
\makeatother

\newcommand*{\myexternaldocument}[1]{%
    \externaldocument{#1}%
    \addFileDependency{#1.tex}%
    \addFileDependency{#1.aux}%
}

\usepackage{xr} 
\myexternaldocument{Supplementary_materials}

\usepackage{pifont} 

\begin{document}

\def\spacingset#1{\renewcommand{\baselinestretch}%
{#1}\small\normalsize} \spacingset{1}


\if0\blind
{
\title{\bf Monitoring Covariance in Multichannel Profiles via Functional Graphical Models}
\author[]{Christian Capezza}
\author[]{Davide Forcina}
\author[]{Antonio Lepore\thanks{Corresponding author. e-mail: \texttt{antonio.lepore@unina.it}}}
\author[]{Biagio Palumbo}

\affil[]{Department of Industrial Engineering, University of Naples Federico II, Piazzale Tecchio 80, 80125, Naples, Italy}

\setcounter{Maxaffil}{0}
\renewcommand\Affilfont{\itshape\small}
\date{}
\maketitle

} \fi

\if1\blind
{
  \begin{center}
    {\LARGE\bf Monitoring Covariance in Multichannel Profiles via Functional Graphical Models}
\end{center}
} \fi

\begin{abstract}
Most statistical process monitoring methods for multichannel profiles focus solely on the mean and are almost ineffective when changes involve the covariance structure.
Although it is known to be crucial, covariance monitoring requires estimating a much larger number of parameters, which may shift in a subtle and sparse fashion. 
That is, an out-of-control (OC) state may manifest with small deviations and affect only a very limited subset of these parameters.
To address these difficulties, we propose a multichannel profile covariance (MPC) control chart based on functional graphical models that provide an interpretable representation of conditional dependencies between profiles.
A nonparametric combination of the likelihood-ratio tests corresponding to different sparsity levels is then used to draw an overall inference and signal whether an OC state may have occurred. 
Between-profile relationships that are likely to have shifted are naturally identified at no additional computational cost.
An extensive Monte Carlo simulation study compares the MPC control chart with state-of-the-art competitors, and a case study on monitoring multichannel temperature profiles in a roasting machine illustrates its practical applicability. 
\end{abstract}

\noindent%


\spacingset{1.45} 

\section{Introduction}
\label{sec_intro}
Recent advances in sensing technologies in modern manufacturing systems have enabled the collection of large amounts of data, which can be used for statistical process monitoring (SPM) and fault diagnosis. Often, these data vary with time or space and are best represented by functional variables or profiles \citep{ramsay,kokoszka}, thereby stimulating growing interest in profile monitoring \citep{noorossana2011statistical}. 
Profile monitoring and SPM in general are commonly implemented in two phases. The first phase (Phase I) is essentially concerned with identifying a clean dataset to be assumed as representative of the in-control (IC) state of the process, hereinafter referred to as the Phase I sample, through retrospective monitoring of an initial dataset drawn from the process; whereas the second phase (Phase II) is concerned with prospective (on-line) monitoring of new observations \citep{qiu2013introduction}, hereinafter referred to as Phase II observations.
The performance of SPM methods is usually measured by the average run length ($ARL$), i.e., the average number of observations required to signal an OC state with respect to a specified shift. The ARL corresponding to no shift is generally referred to as IC $ARL$ or $ARL_0$.

In many practical situations, product quality is characterized by profiles collected from multiple sensors, referred to as channels, that simultaneously measure the same or similar process variables. Multichannel profile data are common in real-world applications, 
as shown by \cite{paynabar2013monitoring, capezza2024robust, li2018pairwise}.
In this setting, several methods have been proposed in the literature with the aim of monitoring the process mean alone. \cite{paynabar2016change} introduced a  Phase I monitoring framework by incorporating a multidimensional functional
principal component analysis (PCA) into a changepoint
model; \cite{ren2019phase} integrated a multichannel functional PCA (MFPCA) with an EWMA scheme for Phase II monitoring;
\cite{wang2018thresholded} proposed a thresholded multivariate PCA method to select significant features in multichannel profiles under the OC state; 
\cite{zhang2018weakly} developed a sparse multichannel functional PCA for weakly correlated multichannel profiles with sparse mean changes.
Other relevant contributions include \cite{paynabar2013monitoring,grasso2014profile,zhang2018multiple,li2018pairwise}.

Unexpectedly, none of these methods addressed monitoring changes in the process covariance structure, which was assumed to be stable over time. However, in practice, this is often not guaranteed, as the state of the process may depend heavily on the dispersion and interactions among quality variables, even when there is no mean shift.
To illustrate this, let us consider the monitoring of multichannel temperature profiles collected from a roasting machine, which will also be examined in Section \ref{sec_real}.
The roasting machine is equipped with five chambers, each containing three temperature sensors positioned at different locations to capture distinct thermal behaviors during the process. A total of 15 temperature profiles are then recorded for every roasting cycle, each lasting 1 hour.
The quality of the final product is known to strongly depend on between-profile relationships, even when the mean remains stable, because they reflect the overall dynamics of the process under study. Such changes may reveal altered heat-transfer dynamics, non-uniform air circulation, or partial failures in the heating elements.
Under IC conditions, temperature profiles collected within the same chamber are strongly correlated because they describe the same physical phenomenon in the same environment, although they were measured at slightly different positions, and thus constitute a valid example of multichannel profile data.
Therefore, a change in the correlation structure due to an unexpected weakening or strengthening provides an early indication of a potential shift in the underlying process dynamics.

This principle extends beyond industrial processes. As an example, in economic and financial markets, intraday price data, yield and term structure data, and intraday volatility data also take the form of profiles \citep{horvath2014testing}. Monitoring the between-profile relationships in such data is essential for proper asset allocation, as shifts in dependency structures can significantly impact portfolio management \citep{avanesov2018change, keshavarz2020sequential}.
Similarly, in volcano activity monitoring \citep{zhao2020distributed}, seismic sensors deployed at multiple locations generate periodic seismic profiles. Although the mean level of seismic activity may remain unchanged, shifts in between-profile relationships should not be overlooked in monitoring volcanic dynamics.

In this framework, functional graphical models are receiving increasing attention because they offer an interpretable representation of the conditional relationships among profiles. In a functional graphical model, each node represents a functional variable, and each edge represents the conditional dependence between two nodes. That is, two nodes are connected if and only if they are dependent conditional on all the others.
Under the assumption of a multivariate Gaussian process, several methods have been proposed in the literature to estimate functional graphical models.
\cite{qiao2019functional} proposed the functional graphical lasso, which extends, to the functional setting, the graphical lasso of \cite{friedman2008sparse}, proposed to estimate graphical models for multivariate scalar data.
The authors reduced the dimensionality of the problem to a finite set of principal component scores via a truncated Karhunen-Loève expansion. Then, the conditional dependencies of the graph are recovered from the nonzero blocks of the precision matrix of the scores.
\cite{zhao2024high} extended the neighborhood selection approach for scalar data of \cite{meinshausen2006high} to the functional setting, using a set of penalized function-on-function regressions.
\cite{zhu2016bayesian} proposed a framework for Bayesian inference of decomposable functional graphs. \cite{zapata2022partial} proposed a new notion of separability for the covariance operator of multivariate functional data.

However, functional graphical model potential has received much less attention in SPM. The work of \cite{wu2022monitoring} is among the few attempts to monitor between-profile relationships in heterogeneous multichannel profiles.
The authors proposed a modeling approach based on heterogeneous graphical models to simultaneously track both differences and similarities among the graphical networks. Then, a Shewhart-type control chart is constructed to monitor the model residuals, although it is well known that such a control chart is not the most appropriate for detecting small and persistent shifts. 
Another drawback of the \cite{wu2022monitoring} approach is the inability to identify changes in process covariance that affect only a small subset of between-profile relationships, which are far more numerous than the variables themselves, with magnitudes and patterns that are unknown in advance.
The idea is that the sparsity of the precision matrix can be used to improve the performance of the control charting mechanism \citep{yeh2012monitoring}.

Along this line, in the multivariate scalar case,
\cite{li2013monitoring} proposed a Shewhart-type control chart based on penalized likelihood estimation of the precision matrix with $\ell_1$ penalty, while \cite{maboudou2013lasso} introduced a similar framework when the estimated sample covariance matrix is singular. \cite{abdella2019adaptive} criticized the shortcomings of penalized likelihood ratio methods and proposed a control chart based on the adaptive thresholding method of \cite{cai2011adaptive}.
\cite{yeh2012monitoring} integrated the penalized likelihood estimation of \cite{li2013monitoring} with the multivariate exponentially weighted moving covariance (MEWMC) framework of \cite{hawkins2008multivariate}. 
See also \cite{maboudou2013lasso}.
\citet{shen2014new} proposed using the classical $L_2$-norm-based test alongside a maximum-norm-based test to overcome the limitations associated with the choice of penalty parameter.
However, none of the aforementioned methods can handle multichannel profile data, and selecting the penalty parameter when monitoring individual observations remains a critical issue, as the performance of penalized likelihood methods strongly depends on this choice. 
Nevertheless, these approaches are primarily designed for Phase II monitoring, assuming that the IC parameters are known or can be perfectly estimated without uncertainty. This assumption is often unrealistic, as uncertainty in parameter estimation can significantly affect the performance of SPM methods \citep{jensen2006effects}.

Another work that incidentally mentions monitoring the covariance of multichannel profile data is that of \cite{ren2019phase}, who envisions an extension of their method based on the MEWMC of \cite{hawkins2008multivariate} and proposes an appropriate likelihood ratio test (LRT) statistic. 
Although scores are assumed to follow a specified covariance structure under IC conditions, the LRT statistic is derived under the assumption of an identity covariance matrix, and no precise guidelines are provided for calculating the control limit with estimated parameters. 
Nevertheless, the method of \cite{ren2019phase}  is not adaptive to the various shift patterns commonly encountered in the multivariate scalar covariance monitoring literature.

The objective of this article is to propose a new SPM method, named the multichannel profile covariance (MPC) control chart, for online monitoring of the covariance of multichannel profiles, capable of identifying small persistent shifts of varying sizes, thereby enabling more timely detection.
The MPC control chart uses MFPCA to extract informative features from the profiles and leverages the interpretable framework of functional graphical models to embed the between-profile relationships into the precision matrix. It is designed to localize the most significant shifted relationships and combine multiple constrained LRTs using the nonparametric combination (NPC)  \citep{pesarin2010finite,pesarin2010permutation,centofanti2025adaptive} to adapt online to unknown shift patterns. When an alarm signal is issued, the proposed framework can also provide a post-signal diagnostic procedure without additional computation, i.e., interpret which specific profile relationships have shifted. 

The remainder of the article is organized as follows. Section \ref{methodology} illustrates the MPC control chart framework. Section \ref{sec_sim} compares the performance of the proposed method with respect to state-of-the-art methods in the literature by means of a Monte Carlo simulation study. In Section \ref{sec_real}, a case study on the monitoring of multichannel temperature profiles in a roasting machine illustrates the practical applicability of the proposed method. Section \ref{sec_conclusions} concludes the article. Supplementary Materials include additional simulation results, details of the data-generation process used in the simulation study, and the algorithms presented in Section \ref{methodology}.
All calculations and plots were obtained using the programming language R \citep{r2021}.

\section{Methodology}
\label{methodology}
The proposed MPC control chart is described in four subsections. Section 
\ref{modeling} presents how the relationship between profiles is modeled using MFPCA and functional graphical models.
In Section \ref{monitoring}, the monitoring procedure based on the MPC statistic and the nonparametric combination of constrained LRTs is described, while a post-signal diagnostic strategy is introduced in Section \ref{post-signal}. 
Section \ref{implementation} provides implementation details.

\subsection{Modeling Multichannel Profile Relationships}
\label{modeling}
Let us assume that $N$ independent and identically distributed observations of the $p$-channel profiles $\boldsymbol{X}(t) = \left(X_1(t),\dots, X_p(t)\right)^T$ are available, with $X_j(t)$, $j=1,\dots,p$, having values in $L^2(\mathcal{T})$, that is, the Hilbert space of square-integrable functions defined in the compact set $\mathcal{T} \subseteq \mathbb{R}$.
Moreover, let us assume that the multichannel profiles follow the model
\begin{equation*}
    \boldsymbol{X}(t) = \boldsymbol{\mu}(t) + \boldsymbol{Y}(t), \quad t \in \mathcal{T},
\end{equation*}
where $\boldsymbol{\mu}(t) = \left(\mu_1(t),\dots,\mu_p(t)\right)^T$ denotes the $p$-dimensional mean function and $\boldsymbol{Y}(t) = \left(Y_1(t),\dots,Y_p(t)\right)^T$ is the stochastic error with $\E\left(\boldsymbol{Y}(t)\right) = \boldsymbol{0}$. Without loss of generality, we let $t \in \mathcal{T} = \left[0,1\right]$, and the mean function $\boldsymbol{\mu}(t)$ to be zero.
Using MFPCA \citep{paynabar2016change}, $\boldsymbol{Y}(t)$ can be represented by a set of orthonormal basis functions as follows
\begin{equation}
\label{KL exp}
    \boldsymbol{Y}(t) = \sum_{k = 1}^\infty \boldsymbol{\xi}_k v_k(t), \quad t \in [0,1],
\end{equation}
where $\boldsymbol{\xi}_k = \left(\xi_{k1},\dots, \xi_{kp}\right)^T = \int_0^1 \boldsymbol{Y}(t)v_k(t)dt$ are zero-mean $p$-variate principal component scores with covariance matrix $\boldsymbol{\Omega}_{0k} = \E\left(\boldsymbol{\xi}_k\boldsymbol{\xi}_k^T \right)$.
Under the multichannel profiles assumption, the elements $Y_j(t)$, $j = 1,\dots,p$, of $\boldsymbol{Y}(t)$ exhibit similar patterns, that is, they share a common set of eigenfunctions $v_k(t)$, $k = 1,2,\dots$. To obtain them, it is sufficient to investigate the overall covariance function defined as
\begin{equation}
\label{pooled covariance}
    c(s,t) = \sum_{j= 1}^p \E\left(Y_j(s)Y_j(t) \right), \quad  s,t \in [0,1],
\end{equation}
and perform standard PCA, as described in Section \ref{implementation}. Then, the corresponding eigenvalues can be obtained as $\lambda_k = \sum_{j = 1}^p \E\left(\xi_{kj}^2\right)$. Equation \eqref{pooled covariance} evidently pools the functional covariance of each channel to measure the \textit{overall covariance} of the vector-valued random function $\boldsymbol{Y}(\cdot)$ for $s,t \in [0,1]$.

In practice, to deal with the infinite dimensionality of $\boldsymbol{Y}(t)$, it is generally assumed that a few eigenvalues and eigenfunctions are able to capture most of the variation in a multichannel profile sample. Thus, $\boldsymbol{Y}(t)$ can be approximated by truncating the summation in Equation \eqref{KL exp}, that is
\begin{equation}
    \label{truncated KL}
        \boldsymbol{Y}(t) = \sum_{k = 1}^K \boldsymbol{\xi}_k v_k(t), \quad t \in [0,1],
\end{equation}
where $K$ is chosen to account for a wanted fraction of total variance explained (FVE).

At this point, it is convenient to introduce a $pK \times pK$ permutation matrix $\boldsymbol{P}$ that allows to arrange the scores $\boldsymbol{\xi} = \boldsymbol{P} \boldsymbol{\eta}$, where $\boldsymbol{\eta} =(\boldsymbol{\xi}_1^T, \dots, \boldsymbol{\xi}_K^T)^T$ in such a way that the first group of $K$ scores corresponds to the first variable, the second group to the second variable, and so on.
As an example, if $p=3$ and $K=2$,
\begin{spacing}{1}
$$
\boldsymbol{P} = 
\begin{pmatrix}
    1 & 0 & 0 & 0 & 0 & 0 \\
    0 & 0 & 0 & 1 & 0 & 0 \\
    0 & 1 & 0 & 0 & 0 & 0 \\
    0 & 0 & 0 & 0 & 1 & 0 \\
    0 & 0 & 1 & 0 & 0 & 0 \\
    0 & 0 & 0 & 0 & 0 & 1
\end{pmatrix},
$$
\vspace{0.5pt}
\end{spacing}
\noindent then, the scores are $\boldsymbol{\xi}=\boldsymbol{P}\boldsymbol{\eta}=(\xi_{11},\xi_{21},\xi_{12}, \xi_{22},\xi_{13},\xi_{23})^T$, with $\boldsymbol{\eta}=(\xi_{11},\xi_{12},\xi_{13},\xi_{21},\xi_{22},\xi_{23})^T$.

Under the assumption that $\boldsymbol{X}(t)$ is a realization of a multivariate Gaussian process, the scores are normally distributed, and the problem of investigating the between-profile relationships can be reduced to calculating the corresponding $pK \times pK$ covariance matrix.
Specifically, as shown by \cite{qiao2019functional}, the conditional dependence structure among $p$ random functions, represented by a graphical network with vertices accounting for the functional variables and edges for the conditional relationships, can be recovered by estimating the precision matrix $\boldsymbol{\Theta}_0=({\boldsymbol{\Sigma}_0})^{-1}$, where $\boldsymbol{\Sigma}_0 = \boldsymbol{P} \boldsymbol{\Omega}_0 {\boldsymbol{P}}^T$ is the covariance matrix of $\boldsymbol{\xi}$ with $\boldsymbol{\Omega}_0$ built as a block-diagonal matrix with $p \times p$ diagonal blocks $\boldsymbol{\Omega}_{0k}$ \citep{paynabar2016change}.

\noindent In particular, supposing that the $X_j(t)$'s, $j = 1,\dots,p$, belong to a graph $G = \left(V,E\right)$ with node set $V = \left\{1,\dots,p\right\}$, the edge set $E$ corresponds to
\begin{equation*}
      E = \{(j,l) : \|\boldsymbol{\Theta}_{0jl}\|_{\text{F}} \neq 0, (j,l) \in V^2, j\neq l \},
\end{equation*}
where $\boldsymbol{\Theta}_{0jl}$ is the $K\times K$ matrix corresponding to the $(j,l)$-th submatrix of $\boldsymbol{\Theta}_0$ and $\|\cdot\|_F$ denotes the Frobenius norm.

Given $N$ independent realizations $\boldsymbol{X}_i(t)$ of $\boldsymbol{X}(t)$, with $p$-dimensional score vectors $\boldsymbol{\xi}_{i,k}$, $i = 1,\dots,N$, $k = 1,\dots,K$,
the estimate $\widehat{\boldsymbol{\Theta}}_0$ of $\boldsymbol{\Theta}_0$ is obtained by solving the following penalized likelihood problem with adaptive lasso penalty
\begin{equation}
\label{afglasso}
      \widehat{\boldsymbol{\Theta}}_0 = \argmax_{\boldsymbol{\Theta}}\biggl\{\log\lvert\boldsymbol{\Theta}\lvert - \operatorname{tr}\left(\widehat{\boldsymbol{\Sigma}}_0\boldsymbol{\Theta} \right)  - \lambda\sum_{j \leq l}\frac{\|\boldsymbol{\Theta}_{jl}\|_{\text{F}}}{\|\Tilde{\boldsymbol{\Theta}}_{0jl}\|_{\text{F}}} \biggl\},
\end{equation}
where, $\widehat{\boldsymbol{\Sigma}}_0 = \boldsymbol{P} \widehat{\boldsymbol{\Omega}}_0\boldsymbol{P}^T$, with $\widehat{\boldsymbol{\Omega}}_0$ obtained by building the block-diagonal matrix with diagonal blocks $\widehat{\boldsymbol{\Omega}}_{0k} = \frac{1}{N}\sum_{i = 1}^N \left(\boldsymbol{\xi}_{i,k}\boldsymbol{\xi}_{i,k}^T \right)$ for each $k = 1,\dots,K$, and $\lambda$ is a nonnegative tuning parameter to be chosen in cross-validation \citep{wu2022monitoring}.
The problem \eqref{afglasso} can be solved by using the alternating direction method of multipliers (ADMM) \citep{boyd2011distributed} as detailed in the Supplementary Materials B.
The adaptive lasso weights involve a nonsparse initial estimator $\Tilde{\boldsymbol{\Theta}}_0$, which is obtained using the Ridge precision matrix estimator of \cite{van2016ridge}. The Ridge estimator replaces the penalty $\ell_1$ in Equation \eqref{afglasso} with a penalty $\ell_2$ of the form
\begin{equation*}
    \frac{\gamma}{2}\operatorname{tr}\left[(\boldsymbol{\Theta}-\boldsymbol{T})^T(\boldsymbol{\Theta}-\boldsymbol{T})\right],
\end{equation*}
with $\gamma$ a nonnegative tuning parameter to be chosen by cross-validation, and $\boldsymbol{T}$ a target matrix chosen here to be the zero matrix. \cite{van2016ridge} provided a closed-form expression for this estimator and established its asymptotic consistency, making it a suitable estimator to construct adaptive lasso weights.

The adaptive lasso penalty in Equation \eqref{afglasso} used in this article differs from the one proposed in \cite{qiao2019functional}, because different amounts of shrinkage are applied to the different blocks of the precision matrix. 
In the multivariate scalar setting, the adaptive lasso has been shown to produce more sparse estimates of the precision matrix compared to the standard lasso. Moreover, the resulting estimator is consistent and asymptotically unbiased \citep{fan2009network}.
However, both the lasso and the adaptive lasso yield biased estimators in finite samples. This is a critical issue because even a small bias in the estimators of the IC parameters can lead to failure to detect a small shift in the bias direction, similar to what occurs in biased statistical tests \citep{qiu2003nonparametric}.
Unfortunately, numerical investigations indicate that the adaptive lasso penalty alone does not sufficiently mitigate bias in finite samples. In fact, when the shift has a magnitude comparable to the bias and is in the bias direction, i.e., it reduces the off-diagonal blocks of $\boldsymbol{\Theta}_0$ to zero, the OC ARL exceeds the corresponding IC ARL.
To address this limitation, we adopt the de-sparsified precision matrix estimator proposed by \cite{jankova2015confidence} and defined as
\begin{equation}
    \label{desparse}
    \widehat{\boldsymbol{\Theta}}_0^* = 2\widehat{\boldsymbol{\Theta}}_0 - \widehat{\boldsymbol{\Theta}}_0\widehat{\boldsymbol{\Sigma}}_0\widehat{\boldsymbol{\Theta}}_0.
\end{equation}
The authors showed that the estimator is unbiased under sufficient sparsity (see Lemma 1 in \cite{jankova2015confidence}), and, in combination with the adaptive lasso penalty, we found it effective in reducing the bias of finite-sample estimates.

\subsection{Monitoring Procedure}
\label{monitoring}
Let us assume that a set of functional principal component scores $\boldsymbol{\xi}_i = \boldsymbol{P} \boldsymbol{\eta}_i$, where $\boldsymbol{\eta}_i = (\boldsymbol{\xi}_{i,1}, \dots \boldsymbol{\xi}_{i,K})^T$ corresponds to the multichannel profile $\boldsymbol{X}_i(t)$, $i = 1,\dots,N$, is available for a Phase I sample. Under the multivariate Gaussian process assumption, the scores $\boldsymbol{\xi}_i$ are independent and identically distributed as $N \sim \left(\boldsymbol{0},\boldsymbol{\Theta}_0^{-1}\right)$, where $\boldsymbol{\Theta}_0$ is the IC precision matrix to be estimated as described in Section \ref{modeling}.
Let $\boldsymbol{X}_n(t)$, $n = N+1, N+2,\dots$, be a new vector of multichannel profiles collected online at a point in time $n$.
The corresponding scores $\boldsymbol{\xi}_n = \boldsymbol{P} \boldsymbol{\eta}_n$, where $\boldsymbol{\eta}_n = (\boldsymbol{\xi}_{n,1}, \dots \boldsymbol{\xi}_{n,K})^T$, are obtained
by projecting $\boldsymbol{X}_n(t)$ onto the space spanned by the eigenfunctions $v_k(t)$ as follows
\begin{equation*}
    \boldsymbol{\xi}_{n,k} = \int_0^1 \boldsymbol{X}_n(t)v_k(t)dt, \quad k=1,\dots,K, \quad n = N+1,N+2,\dots.
\end{equation*}
Let $\boldsymbol{\Theta}_1$ denote the precision matrix of the newly collected scores $\boldsymbol{\xi}_n$. The goal is to test the null hypothesis $H_0$ (IC process) versus the alternative hypothesis $H_1$ (OC process)
\begin{equation*}
    H_0: \boldsymbol{\Theta}_1 = \boldsymbol{\Theta}_0, \quad H_1: \boldsymbol{\Theta}_1 \neq \boldsymbol{\Theta}_0,
\end{equation*}
which corresponds to the detection of changes in the covariance of the scores, given that $\boldsymbol{\Theta}_0 = ({\boldsymbol{\Sigma}_0})^{-1}$.
To test the above hypothesis, as in \cite{hawkins2008multivariate}, we accumulate the historical values of $\boldsymbol{\xi}_{n,k}$ through the MEWMC statistic
\begin{equation}
\label{mewmc}
    \boldsymbol{S}_{n,k} = \left(1 - \rho\right)\boldsymbol{S}_{n-1,k}+ \rho\boldsymbol{\xi}_{n,k}\boldsymbol{\xi}_{n,k}^T, \quad k=1,\dots,K,  \quad n = N+1,N+2,\dots,
\end{equation}
where $\boldsymbol{S}_{0,k} = \widehat{\boldsymbol{\Omega}}_{0,k}$, and $\rho \in \left(0,1\right]$ is the weighting parameter. Then, an estimate of $\boldsymbol{\Theta}_1$ is obtained by solving the following constrained maximum likelihood problem
\begin{align}
\label{constrfgm}
      \widehat{\boldsymbol{\Theta}}_{1s} = \arg\max_{\Theta}  &
\left\{\log\lvert\boldsymbol{\Theta}\rvert - \operatorname{tr}(\boldsymbol{P}\boldsymbol{S}_n\boldsymbol{P}^T\boldsymbol{\Theta})\right\} \\
\text{s.t.} \quad &
\boldsymbol{\Theta}_{jl} = \widehat{\boldsymbol{\Theta}}_{0,jl}, \quad (j,l)\notin I(s),
\nonumber
\end{align}
where $\boldsymbol{S}_n$ is the $pK \times pK$ block diagonal matrix with blocks $\boldsymbol{S}_{n,k}$, and $I(s)$ is a set of $s$ index pairs $(j,l)$, $j,l = 1,\dots,p$. Problem \eqref{constrfgm} using ADMM \citep{boyd2011distributed} as detailed in the Supplementary Materials B.
The estimator $\widehat{\boldsymbol{\Theta}}_{1s}$ is the maximum likelihood estimator of $\boldsymbol{\Theta}_1$, constrained to have elements that do not belong to the set $I(s)$ equal to the corresponding IC counterparts. In other words, only $s$ elements of $\boldsymbol{\Theta}_1$ are estimated, under the assumption that all other entries remain unchanged relative to the IC state.
Ideally, if both the number of shifted elements $s$ and their locations were known, one could simply define the entries of $I(s)$ to be the indices of the shifted elements and thus estimate $\boldsymbol{\Theta}_1$, leaving those entries unconstrained.
 The statistic using a generalized likelihood ratio test (GLRT) can then be constructed as the ratio of the likelihood functions under $H_0$ and $H_1$. Being the log-likelihood function equal, up to a constant, to
\begin{equation}
\label{loglk}
    \ell(\boldsymbol{\Theta}) = \log |\boldsymbol{\Theta}| - \operatorname{tr}\left(\boldsymbol{P}\boldsymbol{S}_n\boldsymbol{P}^T\boldsymbol{\Theta}\right),
\end{equation}
the log-likelihood ratio test statistic can be obtained as
\begin{equation}
    \label{lrt stat}
    \Lambda_s =  \ell(\widehat{\boldsymbol{\Theta}}_{1s}) -  \ell(\widehat{\boldsymbol{\Theta}}_0^*).
\end{equation}
Trivially, larger values of $\Lambda_s$ favor the alternative hypothesis.\\
However, in practice, neither the identity of the shifted elements nor the value of $s$ is known in advance.
Therefore, an identification procedure is required to determine which components of $\boldsymbol{\Theta}_1$ have changed.
For fixed $s$, an exhaustive search would require estimating ${p(p+1)/2}\choose{s}$ candidate precision matrices, and selecting the one that maximizes the log-likelihood ratio statistic in Equation \eqref{lrt stat}. Such a combinatorial search becomes computationally prohibitive in online monitoring, especially in high-dimensional settings, and is therefore impractical.

To overcome this problem, inspired by the covariance estimation procedure of \cite{cai2011adaptive}, also used in monitoring covariance matrices with batched observations \citep{abdella2019adaptive}, we propose to estimate the most significant shifted elements of $\boldsymbol{\Theta}_1$ as follows.
Let $\Tilde{\boldsymbol{\Theta}}_1$ be the nonsparse Ridge estimator of \cite{van2016ridge}, obtained using $\boldsymbol{T} = \widehat{\boldsymbol{\Theta}}_0$ as the target matrix. The choice of the penalty parameter will be addressed in Section \ref{implementation}. We quantify the deviation between $\Tilde{\boldsymbol{\Theta}}_1-\widehat{\boldsymbol{\Theta}}_0$ through the symmetric $p\times p$ matrix $\boldsymbol{D}$, whose elements are the Frobenius norms of the corresponding submatrices
\begin{equation}
\label{diff_frob}
    D_{jl} = \|\Tilde{\boldsymbol{\Theta}}_{1,jl}-\widehat{\boldsymbol{\Theta}}_{0,jl}\|_F \quad j = 1,\dots,p, \quad l = 1,\dots j.
\end{equation}
Let  $p^D_{jl}$ be the  $p$-values associated with $D_{jl}$, for $j = 1, \dots p$, $l = 1\dots,j$, defined as
\begin{equation*}
    p^D_{jl} = \text{Pr}\left(D_{jl} \geq \tilde{D}_{jl} | H_{0} \right), \quad j = 1,\dots p, \quad l = 1,\dots,j,
\end{equation*}
where $\tilde{D}_{jl}$ is the observed value of $D_{jl}$, and denote their
order statistics by
\begin{equation*}
p^D_{(1)} \le p^D_{(2)} \le \cdots \le 
p^D_{\left(\frac{p(p+1)}{2}\right)} .    
\end{equation*}
The $s$ most significant shifted elements are obtained by retaining the indices associated with the $s$ smallest $p$-values
\begin{equation}
\label{indices}
I(s) = 
\left\{
(j,l) : 
p^D_{jl} \in 
\{p^D_{(1)}, \dots, p^D_{(s)}\}
\right\}.    
\end{equation}
Given the selected $I(s)$, the constrained estimator $\widehat{\boldsymbol{\Theta}}_{1s}$ is obtained by solving Equation \eqref{constrfgm} and subsequently plugged into Equation \eqref{lrt stat} to compute $\Lambda_s$.
However, this test statistic is highly sensitive to the choice of $s$, as it directly controls the sparsity of the shift, a problem closely related to the choice of the penalty parameter in penalized likelihood methods \citep{li2013monitoring,yeh2012monitoring}.
As mentioned in the introduction, the use of penalized likelihood methods for monitoring the covariance matrix with a fixed pre-defined penalty parameter has been heavily criticized in the multivariate scalar literature \citep{cai2011adaptive,shen2014new,abdella2019adaptive}, as the shift pattern is usually unknown and could differ from the one for which the selected penalty would be optimal. A large penalty value could, in fact, favor the detection of changes characterized by sparse shift patterns, whereas a small value might be more effective in identifying changes with dense shift patterns.
In order to be adaptive to various types of changes, we propose combining the partial tests associated with multiple values of the tuning parameter $s$ using a nonparametric combination (NPC) \citep{pesarin2010finite,pesarin2010permutation}.
The NPC is a nonparametric framework that aggregates $ p$-values from dependent or independent partial tests via combining functions and has recently been used to monitor a mean shift in multivariate functional data \citep{centofanti2025adaptive}.
Specifically, the null hypothesis $H_0$ is decomposed into a set of null sub-hypotheses $H_{0s}$, with the corresponding tuning parameter $s \in S$, where $S$ is a set of tuning parameters with cardinality $n_s$, such that $H_0$ is true if all the $H_{0s}$ are simultaneously true. More precisely, the null sub-hypotheses of the relative partial tests and the corresponding alternatives are
\begin{equation*}
        H_{0s}: \boldsymbol{\Theta}_{1s} = \boldsymbol{\Theta}_0, \quad H_{1s}: \boldsymbol{\Theta}_{1s} \neq \boldsymbol{\Theta}_0, \quad s \in S,
\end{equation*}
where $\boldsymbol{\Theta}_{1s}$ is the precision matrix of new scores to be estimated at given $s$.
Partial tests are combined on a common scale in an overall monitoring statistic $\Lambda$ considering the corresponding $p$-values $p_s$, defined as
\begin{equation*}
    p_s = \text{Pr}\left(\Lambda_s \geq \tilde{\Lambda}_s | H_{0s} \right), \quad s \in S,
\end{equation*}
where $\tilde{\Lambda}_s$ denotes the observed value of $\Lambda_s$.
While several combination approaches are possible to obtain the overall monitoring statistic $\Lambda$, we propose the \textit{Fisher omnibus} \citep{fisher1970statistical} combining function, which represents a popular choice, through which we obtain
\begin{equation}
\label{fisher omnibus}
    \Lambda = -2\sum_{s \in S}\log p_s.
\end{equation}
The MPC control chart then signals when $\Lambda$ exceeds a control limit $h$, chosen as the smallest value achieving an IC $ARL$ larger than or equal to a pre-specified value, denoted by $ARL_0$. More details on the choice of $h$ and the implementation of the proposed method are given in Section \ref{implementation}.\\
It should be noted that the proposed approach differs substantially from both penalized likelihood formulations and the adaptive procedure of \citet{abdella2019adaptive} in the following main aspects.
Different from the latter, the proposed method is specifically designed for multichannel profile data.
Secondly, penalized likelihood methods, as well as the adaptive thresholding scheme proposed by \citet{abdella2019adaptive}, critically depend on the choice of a penalization or thresholding parameter, which strongly influences their detection performance. Although the adaptive thresholding scheme proposed by \cite{abdella2019adaptive} adjusts its thresholds according to the estimated variability of the covariance elements, it cannot avoid user-specified tuning to directly govern the level of sparsity in the estimated shift.
Moreover, the adaptive thresholding scheme is theoretically justified by the asymptotic normality of the covariance estimator, which unfortunately cannot be used in the multichannel profile setting, where the effective sample size is constrained by $\rho$ in the MEWMC statistic \eqref{mewmc}, and the dependency structure is encoded in the blocks of the estimated precision matrix, each containing correlated components.

\subsection{Post-signal Diagnostics}
\label{post-signal}
A key advantage of the proposed MPC control charting scheme is that a diagnostic procedure that can locate a possible change point and identify the between-profile relationship(s) responsible for an OC signal is naturally embedded within the monitoring framework. 
In many industrial settings, this procedure plays a crucial role in the practical identification and elimination of the root cause(s) that may have led to the OC state of the process.

As described in Section \ref{monitoring}, the $p$-values $p^D_{jl}$, $j = 1,\dots p, l = 1,\dots,j$ associated with the deviations $D_{jl}$, $j = 1,\dots p, l = 1,\dots,j$ in Equation \eqref{diff_frob} are already computed at each monitoring step to identify the most significant shifted elements. These $p$-values provide a direct measure of evidence against the null hypothesis for each between-profile relationship and can be used immediately for diagnostic purposes, without requiring any additional computational effort.

To identify any between-profile relationship responsible for the OC signal, we propose using the Benjamini-Hochberg (BH) procedure \citep{benjamini1995controlling}, which controls the false discovery rate (FDR) among the detected shifts, to the $p$-values $p^D_{jl}$, $j = 1,\dots p, l = 1,\dots,j$ corresponding to the triggering of an alarm. The BH procedure addresses the multiple-testing problem inherent in simultaneously examining all pairwise relationships while maintaining control of the proportion of false discoveries.

To identify the change point, a penalized likelihood procedure based on the $\ell_2$ penalty is adopted.
In particular, suppose that the proposed MPC control chart triggers a signal at time step $m$. An estimator $\widehat{\tau}$ of the change point $\tau$ is obtained by considering the following log-likelihood function at each candidate change point $u$,
\begin{align}
\label{llcpd}
    \ell^{cp}(u,m,\boldsymbol{\Theta}_u) = &N_{b,u}\left( \log |\widehat{\boldsymbol{\Theta}}_0^*| - \operatorname{tr}\left(\boldsymbol{P}\boldsymbol{S}_{b,u}\boldsymbol{P}^T\widehat{\boldsymbol{\Theta}}_0^* \right)\right)
    + \\
    & N_{a,u}\left(\log |\boldsymbol{\Theta}_u| - \operatorname{tr}\left(\boldsymbol{P}\boldsymbol{S}_{a,u}\boldsymbol{P}^T\boldsymbol{\Theta}_{u} \right)\right),
    \nonumber
\end{align}
where $N_{b,u}$ and $N_{a,u}$ are the number of observations before and after $u$, respectively. Analogously, $\boldsymbol{S}_{b,u}$ and $\boldsymbol{S}_{a,u}$ are the sample covariance matrices of the scores collected before and after $u$, and $\boldsymbol{\Theta}_{u}$ is the precision matrix of the scores collected after $u$.
An estimate $\tilde{\boldsymbol{\Theta}}_{u}$ of $\boldsymbol{\Theta}_{u}$ can be obtained by using the Ridge estimator of \cite{van2016ridge} with target matrix $\boldsymbol{T} = \widehat{\boldsymbol{\Theta}}_0$.
Thus, the change point can be finally estimated as
\begin{equation}
\label{estcp}
    \widehat{\tau} = \argmax_{1\leq u \leq m}  \ell^{cp}(u,m,\tilde{\boldsymbol{\Theta}}_u).
\end{equation}

\subsection{Implementation Details}
\label{implementation}
In Section \ref{modeling}, it is assumed that the IC mean function $\boldsymbol{\mu}(t)$ and the eigenfunctions $v_k(t)$ are known; however, they have to be estimated from a set of IC data.
In practice, profile observations $\boldsymbol{X}_i(t)$, $i = 1,\dots, N$, are often measured in an ordered and dense grid of points $t_{ih}$, $h = 1,\dots,n_i$, within an interval of finite length \citep{ramsay}.
However, in many cases, $n_i$ is fixed and $t_{ih}$ depends only on $h$, that is, $t_{h}$, $h = 1,\dots,n$. Then, an estimator $\widehat{\boldsymbol{\mu}}(t)$ of the mean function $\boldsymbol{\mu}(t)$ can be easily obtained by its sample version for each grid point, and the MFPCA can be solved by applying traditional PCA to the $n\times n$ sample covariance matrix $\widehat{\boldsymbol{C}}$ with elements
\begin{equation*}
    \widehat{C}_{hl} = \frac{1}{N}\sum_{i = 1}^N \sum_{j = 1}^p \left\{X_{ij}(t_h)-\widehat{\mu}_j(t_h) \right\}\left\{X_{ij}(t_l)-\widehat{\mu}_j(t_l)\right\}, \quad h,l = 1,\dots n.
\end{equation*}
Then, the entries $\sigma_{0kjq}$ of $\boldsymbol{\Omega}_{0k}$ are estimated as
\begin{equation*}
    \widehat{\sigma}_{0kjq} = \frac{1}{N}\sum_{i = 1}^N \left[\sum_{h = 1}^n \left\{X_{ij}(t_h)-\widehat{\mu}_j(t_h) \right\}\widehat{v}_k(t_h)
    \sum_{h = 1}^n \left\{X_{iq}(t_h)-\widehat{\mu}_q(t_h) \right\}\widehat{v}_k(t_h)\right],
\end{equation*}
where $\widehat{v}_k(t_h)$ are the estimated eigenfunctions of $\widehat{\boldsymbol{C}}$ evaluated at $t_h$, obtained at the previous step. If the sampling grid is sparse or the grid points are unevenly spaced, individual profile data can first be smoothed using appropriate smoothing techniques, and then the above procedure can be applied to an equally spaced grid.

The set of tuning parameters $S$ can be selected based on the number of between-profile relationships that are expected to shift in practice. In particular, let 
$s^*$ denote the true number of shifted blocks. Values of $s\in S$ that are close to $s^*$ tend to enhance the detection capability of the proposed method, whereas values that deviate substantially from $s^*$ may reduce performance when the corresponding $p$-values are aggregated as in Equation \eqref{fisher omnibus}.
Therefore, when prior domain knowledge or engineering expertise indicates the plausible sparsity level of the shift, we recommend selecting values of $s$ within that range. In the absence of such knowledge, a practical strategy is to consider a moderately sized grid of candidate values. 
In our experiments, we found that using a grid of size $n_s = 10$, ranging from $1$ to half of the total number of elements, that is, $p(p+1)/4$, provides a good balance between computational cost and detection performance.

The $\ell_2$ penalty parameter $\gamma$ used to calculate $\tilde{\boldsymbol{\Theta}}_1$ in Equation \eqref{diff_frob} balances bias and variance and, consequently, the detection of possibly shifted elements. Specifically, a high $\gamma$ shrinks the elements of $\boldsymbol{\Theta}_1$ toward their corresponding IC values, reducing variance but increasing bias, while a low $\gamma$ yields estimates that are less shrunken but more variable. The optimal $\gamma$ depends on the unknown number, magnitude, and direction of the shift. 
In practice, an effective $\gamma$ can be selected using IC data by estimating $B$ precision matrices $\tilde{\boldsymbol{\Theta}}_b$, $b = 1,\dots, B$, over a grid of values, and choosing $\gamma$ where the decreasing rate of improvement in the mean out-of-sample negative log-likelihood per unit change in $\gamma$ falls below a given threshold; for example, $10^{-3}$. Selecting $\gamma$ in such a way, rather than minimizing the negative log-likelihood, allows a penalty that effectively reduces the variability of IC estimates of $\boldsymbol{\Theta}_1$ without overshrinkage, which can make small shifts in the precision matrix undetectable.

Both the IC distributions of $\Lambda_s$ and $D_{jl}$, $j = 1,\dots,p$, $l = 1,\dots,j$, are usually unknown in practice, and the $p$-values $p_s$ and $p^D_{jl}$, $j = 1,\dots,p$, $l = 1,\dots,j$, are difficult to obtain straightforwardly. To this end, we propose the use of the following estimators \citep{pesarin2010permutation}
\begin{equation}
\label{pval_D}
    \widehat{p}^D_{jl} = \frac{1 + \sum_{i = 1}^{n_{IC}}I\left(D_{i,jl}^{IC} \geq D_{jl} \right)}{1+n_{IC}}, \quad j = 1,\dots,p, \quad l = 1,\dots,j,
\end{equation}
and 
\begin{equation}
\label{pval_s}
    \widehat{p}_s = \frac{1 + \sum_{i = 1}^{n_{IC}}I\left(\Lambda_{i,s}^{IC} \geq \Lambda_s \right)}{1+n_{IC}}, \quad s \in S,
\end{equation}
where $D_{1,jl}^{IC},\dots,D_{n_{IC},jl}^{IC}$ and
$\Lambda_{1,s}^{IC},\dots,\Lambda_{n_{IC},s}^{IC}$ are realizations of $D_{jl}$ and partial test values $\Lambda_s$ calculated by generating $n_{seq}^I$ sequences of length $l_{seq}^I$ by randomly sampling IC data.
Then, for each sequence and observation, the test statistic $\Lambda$ is computed by combining the $p$-values $\widehat{p}_s$, $s \in S$, using Equation \eqref{fisher omnibus}.
Finally, for each sequence, the number of observations acquired up to the first signal of an OC state is stored as a run length ($RL$), and the $ARL$ is calculated as the average of the $RL$ values over the $n_{seq}^I$ sequences. The control limit $h$ is chosen to reach a pre-specified $ARL_0$.

To reduce overfitting issues \citep{kruger2012statistical} in the calculation of the control limit $h$ and the quantities in Equations \eqref{pval_D} and \eqref{pval_s}, we partition the Phase I sample into a \textit{training} and a \textit{tuning} set. The first is used to estimate the MFPCA and the IC precision matrix $\widehat{\boldsymbol{\Theta}}_0$, while the latter is used to calculate the remaining quantities.
Moreover, it is worth noting that the above procedure discards sequences that fail to produce OC signals, and thus requires setting a high $l_{seq}^I$ to minimize the number of discarded iterations. To overcome this, we adopt the approach introduced by \cite{lim2024efficient}, who proposed to retain discarded runs as Type I censored observations, which are then used to estimate the $ARL$. This allows us to set $l_{seq}^I$ to a moderate value while simultaneously providing a good estimate of the control limit $h$, leveraging information from the censored sequences.

In Phase II monitoring, the current observation $\boldsymbol{X}_{new}(t)$ is projected onto the MFPCA model to obtain the scores $\boldsymbol{\xi}_{new,k}$. Then, the MEWMC statistic is calculated using Equation \eqref{mewmc}, and it is used to compute $\tilde{\boldsymbol{\Theta}}_1$ to calculate $\boldsymbol{D}$.
After that, $I(s)$ is obtained through Equation \eqref{indices} and is used to estimate $\boldsymbol{\Theta}_{1s}$ for $s \in S$ by solving Equation \eqref{constrfgm}. Finally, the monitoring statistic $\Lambda$ is computed by combining the $p$-values associated with $\Lambda_s$. An alarm is triggered if $\Lambda > h$ and the post-signal diagnostic procedure of Section \ref{post-signal} is performed to identify the location of the change point and the between-profile relationships responsible for the OC condition.
A detailed algorithm of the proposed method is shown in Algorithm \ref{alg:MPC}.

\begin{algorithm}[ht]
\caption{MPC control chart}
\label{alg:MPC}
\KwIn{IC data $\{ \boldsymbol{X}_i(t) \}_{i=1}^N$, parameters $FVE$, $\rho$, $n_s$, $ARL_0$, $n_{seq}^I$, $l_{seq}^I$, FDR, Phase II data $\boldsymbol{X}_{new}(t)$}
\KwOut{$\widehat{\boldsymbol{\Theta}}_0$, $\widehat{\tau}$, significant shifted elements}

\textbf{Phase I:}\\
\Indp 
Split $\{\boldsymbol{X}_i(t) \}_{i=1}^N$ into \textit{training} set and \textit{tuning} set\;

\textbf{With the \textit{training} set,}\\
\Indp 
Estimate IC parameters $\widehat{\boldsymbol{\mu}}(t)$, $\widehat{v}_k(t)$,
$\widehat{\boldsymbol{\Theta}}_0$ and $\widehat{\boldsymbol{\Theta}}_0^*$\;
\Indm 

\textbf{With the \textit{tuning} set:}\\
\Indp 
Calculate the ridge penalty $\gamma$, the empirical distributions of $D_{jl}$, $j= 1,\dots,p$, $l = 1,\dots,j$ and $\Lambda_s$\;
Generate $n_{seq}^I$ sequences by random sampling $l_{seq}^I$ observations\;
\Indm
\Indm
\For{$i \gets 1$ \KwTo $n_{seq}^I$}{
    \For{$n \gets 1$ \KwTo $l_{seq}^I$}{
        Calculate the MEWMC statistic using Equation \eqref{mewmc}\;
        Calculate the nonsparse estimate $\tilde{\boldsymbol{\Theta}}_1$\;
        Compute the $p$-values of the differences $D_{jl}(i,n) = \|\tilde{\boldsymbol{\Theta}}_{1jl} -\widehat{\boldsymbol{\Theta}}_{0jl}\|$\;
        Obtain $\widehat{\boldsymbol{\Theta}}_{1s}(i,n)$ by solving Equation \eqref{constrfgm} for each $s \in S$\;
        Compute $\Lambda_s(i,n)$ for each $s \in S$\;
        Combine $\widehat{p}_s(i,n)$'s with Equation \eqref{fisher omnibus} to obtain the test statistics $\Lambda(i,n)$\;
    }
}
\Indp
\Indp
Choose the control limit $h$ to achieve the desired $ARL_0$\;
\Indm 
\Indm

\textbf{Phase II:}\\
\Indp 
Compute the MEWMC statistic on new observations using Equation \eqref{mewmc}\;
Calculate the nonsparse estimate $\tilde{\boldsymbol{\Theta}}_1^{new}$\;
Compute the $p$-values of the differences $D_{jl}^{new} = \|\tilde{\boldsymbol{\Theta}}_{1jl}^{new} -\widehat{\boldsymbol{\Theta}}_{0jl}\|$\;
Obtain $\widehat{\boldsymbol{\Theta}}_{1s}^{new}$ by solving Equation \eqref{constrfgm} for each $s \in S$\;
Compute $\Lambda_s^{new}$ for each $s \in S$\;
Combine $\widehat{p}_s^{new}$'s with Equation \eqref{fisher omnibus} to obtain the test statistics $\Lambda^{new}$\;
\Indm 

\If{$\Lambda^{new} > h$}{
    Trigger an alarm and stop monitoring\;
    Perform the post-signal diagnostic procedure of Section \ref{post-signal} to obtain $\widehat{\tau}$ and the shifed elements of $\widehat{\boldsymbol{\Theta}}_0$\;
}
\end{algorithm}

\section{Simulation Study}
\label{sec_sim}
To evaluate the performance of the proposed MPC control chart for detecting an OC state of the process across different scenarios, inspired by \cite{zhao2024high}, IC multichannel profile observations are generated as follows.
Let
\begin{equation*}
    X_{ij}(t) = \boldsymbol{c}_{ij}^T\boldsymbol{\phi}(t), \quad i = 1,\dots,N, \quad j = 1,\dots p, \quad t \in [0,1],
\end{equation*}
be i.i.d. multichannel profile observations, where $\boldsymbol{c}_{ij} \in \mathbb{R}^M$ and $\boldsymbol{\phi}(t)$ is the vector containing the first $M=5$ Fourier basis functions. The vector of coefficients $\left(\boldsymbol{c}_{i1}^T,\dots,\boldsymbol{c}_{ip}^T \right)^T \in \mathbb{R}^{pM}$ follows a multivariate Gaussian distribution with zero mean and covariance matrix $\boldsymbol{\Sigma}_0 = \boldsymbol{\Theta}_0^{-1}$.
Three settings are considered for the precision matrix $\boldsymbol{\Theta}_0$.
\begin{itemize}
    \item \textbf{Model I:} Let $\boldsymbol{A} \in \mathbb{R}^{M\times M}$ be a tridiagonal matrix with $\boldsymbol{A}_{k,k} = 1$, $\boldsymbol{A}_{k,k+1} = \boldsymbol{A}_{k+1,k} = 0.6$ and $\boldsymbol{A}_{k,k+2} = \boldsymbol{A}_{k+2,k} = 0.3$. 
    The precision matrix $\boldsymbol{\Theta}_0$ is generated as a block-banded precision matrix with blocks given as $\boldsymbol{\Theta}_{0j,j} = \boldsymbol{A}$, $\boldsymbol{\Theta}_{0j,j+1} = \boldsymbol{\Theta}_{0j+1,j} = 0.6\boldsymbol{A}$ and $\boldsymbol{\Theta}_{0j,j+2} = \boldsymbol{\Theta}_{0j+2,j} = 0.3\boldsymbol{A}$. All remaining blocks are set to zero.
    \item \textbf{Model II:} The precision matrix $\boldsymbol{\Theta}_0$ is generated as a block diagonal matrix as follows. Profiles are arranged into groups of three. For each group, the corresponding $3M\times 3M$ precision matrix block is generated as in Model I. The off-diagonal blocks are set to zero. If $p$ is not a multiple of three, the remaining variables are assumed to be independent, and the last blocks are $\boldsymbol{\Theta}_{0j,j} = \boldsymbol{A}$.
    \item \textbf{Model III:} $\boldsymbol{\Theta}_0$ is generated as a random block sparse precision matrix with off-diagonal blocks $\boldsymbol{\Theta}_{0j,l} = \boldsymbol{\Theta}_{0l,j} = 0.5\boldsymbol{A}$ with probability 0.2, and $\boldsymbol{0}_M$ otherwise. The diagonal blocks are set as $\boldsymbol{\Theta}_{0j,j} = \delta\boldsymbol{A}$, where $\delta$ is chosen to guarantee the positive definiteness of the precision matrix, i.e., $\boldsymbol{\Theta}_0 \succ 0$.
\end{itemize}

\noindent We fix $N = 2000$, $p = 10,20,30$, while each random function is observed on a grid with 100 equally spaced time points in $[0,1]$. Then, for each time point, the observed values are generated as
\begin{equation*}
    \tilde{X}_{ij}(t) = \sum_{l=1}^M c_{ijl}^T\boldsymbol{\phi}_l(t) + \varepsilon_{ij}(t), \quad i = 1,\dots,N, \quad j = 1,\dots p, \quad t \in [0,1],
\end{equation*}
where $\varepsilon_{ij}(t) \sim N\left(0,0.5^2\right)$.
Although data are observed through noisy discrete values, each component of the generated observations is obtained by the spline smoothing approach with a roughness penalty on the integrated squared second derivative, using $20$ cubic B-splines, with the penalty parameter chosen via the generalized cross-validation criterion (GCV), as commonly described by \cite{ramsay}.

The proposed method is compared with two state-of-the-art methods in the literature, namely, the hierarchical graphical model of \cite{wu2022monitoring}, referred to as HGM, and the multichannel profile monitoring approach of \cite{ren2019phase}, hereinafter referred to as REN. These methods aim to identify shifts in relationships among multiple profiles. 
As noted in the Introduction, \cite{ren2019phase} only briefly mentioned how to monitor the covariance of multichannel profile data. In addition, they gave no specific guidelines for calculating the control limit and derived the monitoring statistic under the assumption of an IC identity covariance matrix of the scores, which is not the case here. This leads to a biased statistical test that, as discussed in Section \ref{modeling}, yields poor detection of shifts near the bias direction. To overcome these limitations, in this article, the monitoring statistic is derived by replacing the IC identity covariance with the correct IC covariance matrix for the scores, $\boldsymbol{\Omega}_{0k}$.
Then, the control limit is computed through the same procedure described in \cite{capezza2025adaptive}.

For each model, four scenarios, each characterized by a different shift pattern, are explored to define the OC precision matrix $\boldsymbol{\Theta}_1$ from the IC precision matrix $\boldsymbol{\Theta}_0$.
In the first scenario (Scenario I), new edges are added to the IC graph. Specifically, we randomly shift zero off-diagonal blocks of $\boldsymbol{\Theta}_0$, such that $\boldsymbol{\Theta}_{0jl} = \boldsymbol{0}$, $j\neq l$, to $\boldsymbol{\Theta}_{1jl} = \delta^\prime\boldsymbol{A}$, where $0\leq \delta^\prime \leq \delta^\prime_{max}$ corresponds to a specific severity level (SL) and $\delta^\prime_{max}$ is a value chosen such that $\boldsymbol{\Theta}_1$ is positive definite and its determinant is close to zero.
In the second scenario (Scenario II), the existing edges of the IC graph are removed as follows. We randomly select non-zero off-diagonal blocks of the IC precision matrix $\boldsymbol{\Theta}_0$, such that $\boldsymbol{\Theta}_{0jl} \neq 0$, $j\neq l$, to $\boldsymbol{\Theta}_{1jl} = (1-\delta^{\prime\prime})\boldsymbol{\Theta}_{0jl}$, where $0\leq \delta^{\prime\prime} \leq 1$ corresponds to a specific SL.
In the third scenario(Scenario III), diagonal blocks of the IC precision matrix are increased by setting $\boldsymbol{\Theta}_{1jj} = (1+\delta^{\prime\prime})\boldsymbol{\Theta}_{0jj}$. This corresponds to a weakening of the relationships involving the $j$-th variable and a decrease in its marginal variance.
Finally, in the fourth scenario, referred to as Scenario IV, diagonal blocks of the IC precision matrix are decreased by setting $\boldsymbol{\Theta}_{1jj} = (1-\delta^{\prime\prime\prime})\boldsymbol{\Theta}_{0jj}$, with $0\leq \delta^{\prime\prime\prime} \leq \delta^{\prime\prime\prime}_{max}$, where $\delta^{\prime\prime\prime}_{max}$ corresponds to a specific SL and being a value chosen such that $\boldsymbol{\Theta}_1$ is positive definite and its determinant is close to zero.
This corresponds to strengthening the relationships involving the $j$-th variable and increasing its marginal variance.
Then, for each scenario, shift patterns are characterized by shifting $n_{el} = 1,3,5,10$ elements of $\boldsymbol{\Theta}_0$, respectively.

Five increasing severity levels $SL \in \{0,1,2,3,4\}$ are explored, each corresponding to a given shift in the precision matrix, where $SL = 0$ corresponds to an IC process. 
For each SL, 50 simulation runs are performed.
According to Section \ref{implementation}, in each run, a training set and a tuning set are
obtained from the simulated IC data by randomly splitting it into $500$ and $1500$ observations, respectively.
The number of functional principal components is chosen so that the fraction of variance explained is at least $95$\%, whereas the MEWMC parameter in Equation \eqref{mewmc} is chosen as $\rho = 0.1$ and $S$ is chosen according to Section \ref{implementation} with cardinality $n_s = 10$. With the tuning set, $n_{seq}^I = 200$ sequences of length $l_{seq}^I = 200$ are generated to estimate the unknown quantities to be used in the monitoring, as detailed in Algorithm \ref{alg:MPC}.
In Phase II, $n_{seq}^{II} = 100$ sequences of length $l_{seq}^{II} = 1000$ are randomly generated, and the performance of the control charts is evaluated by means of the $ARL$ as a function of SL. When the process is IC, the $ARL$ should be as close as possible to $ARL_0$, which is set equal to 100, while it should be as small as possible when the process is OC.
\begin{figure}
    \centering
    \includegraphics[width=1\linewidth]{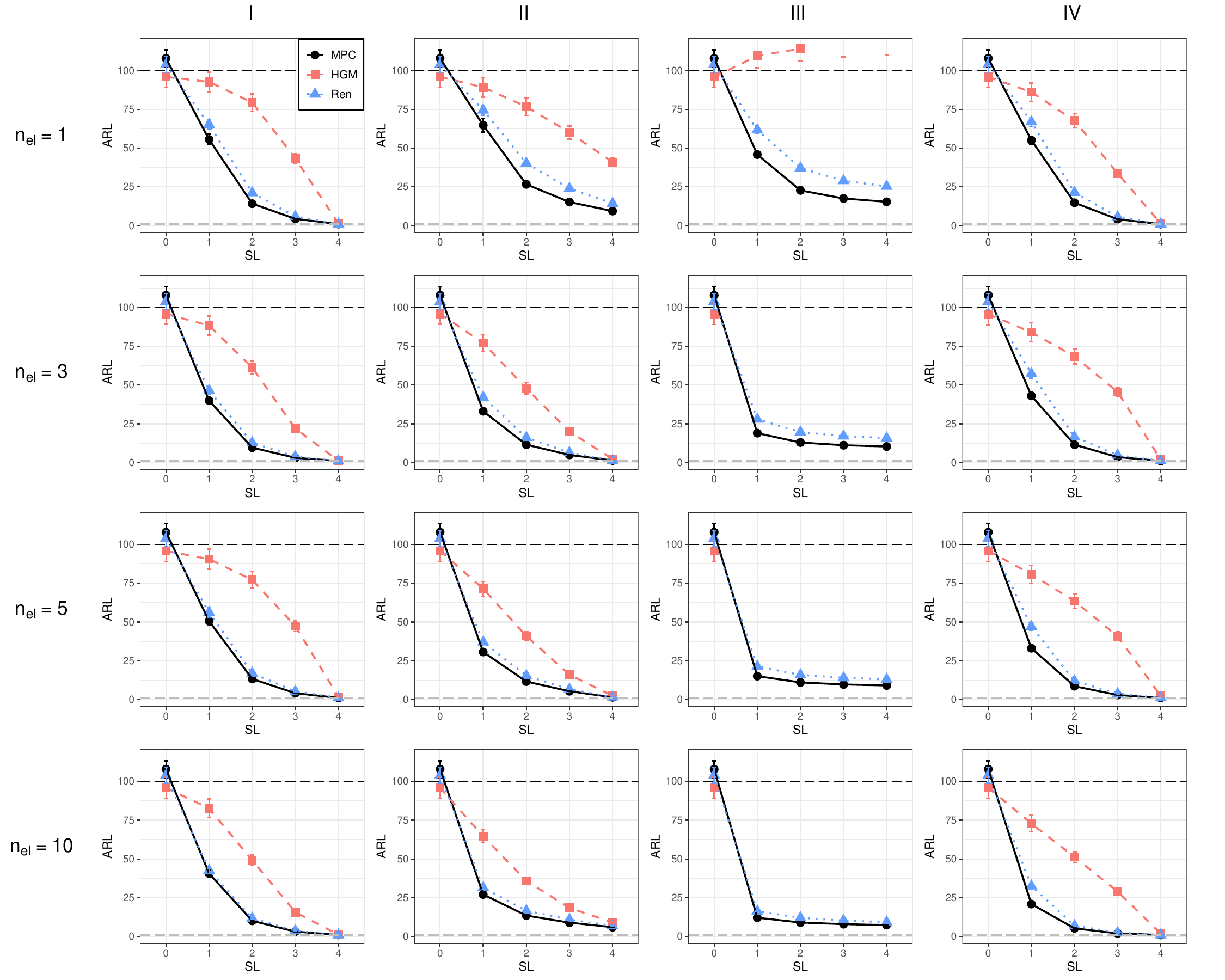}
    \caption{Mean $ARL$ achieved in Phase II by MPC, HGM, and Ren for Model I with $p = 10$ functional variables as a function of SL.}
    \label{fig:p_10_graphI}
\end{figure}

\begin{figure}
    \centering
    \includegraphics[width=1\linewidth]{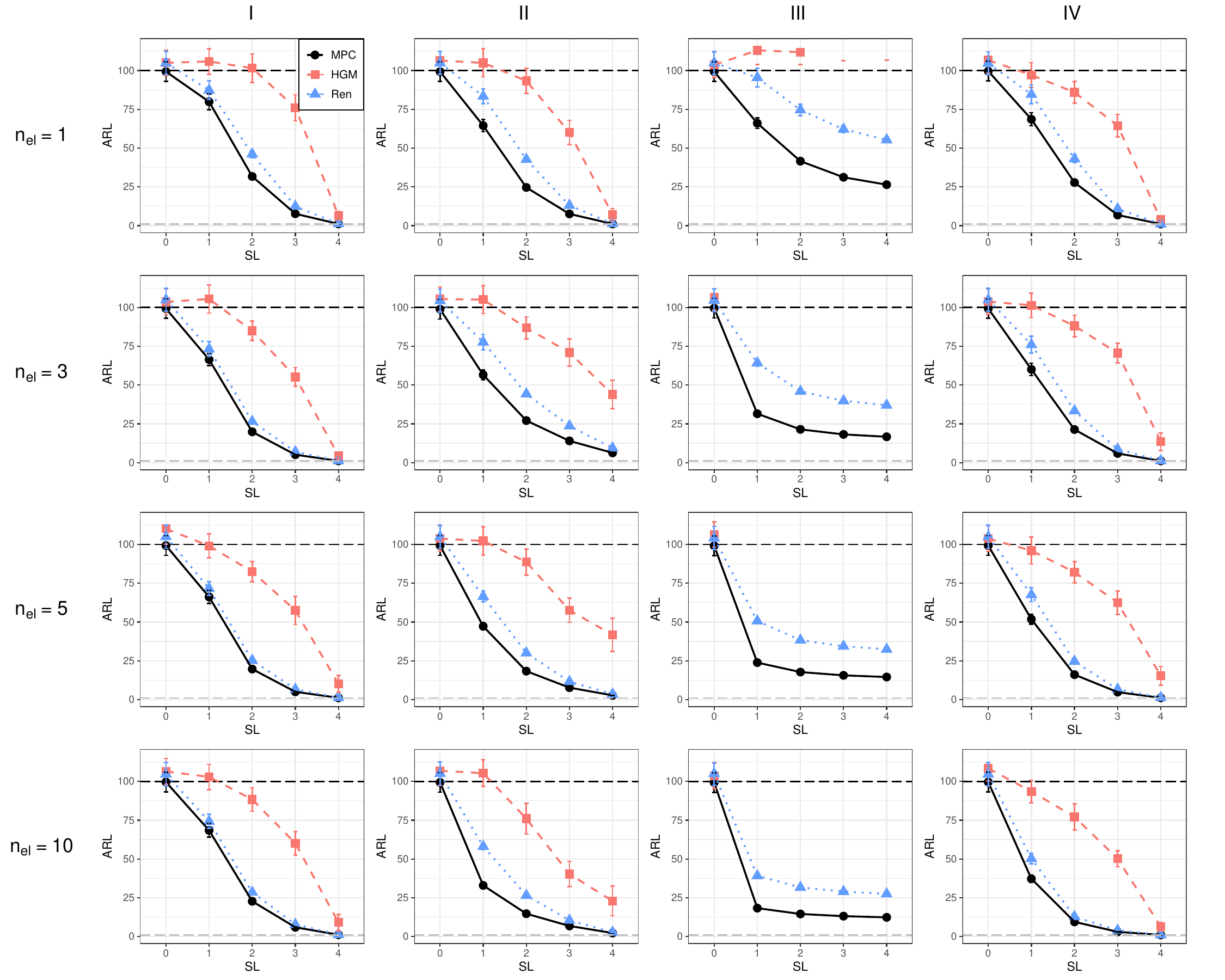}
    \caption{Mean $ARL$ achieved in Phase II by MPC, HGM, and Ren for Model I and $p = 30$ functional variables as a function of SL.}
    \label{fig:p_30_graph_I}
\end{figure}

The simulation results for Model I with $p = 10$ and $p = 30$ are shown in Figure \ref{fig:p_10_graphI} and Figure \ref{fig:p_30_graph_I}, respectively, which depict the mean $ARL$ as a function of SL for each monitoring method and combination of scenarios and shift patterns. 
The proposed MPC control chart clearly outperforms competing methods across all scenarios, particularly for sparser shifts.
HGM performs the worst, as it monitors single observations and does not accumulate historical data.
As expected, REN accumulates historical data as the proposed method and thus performs better than HGM, but struggles to maintain comparable performance to MPC because it does not leverage shift sparsity.
The performance gap between the proposed method and Ren tends to decrease with the number of shifted elements, $n_{el}$, with MPC still the best-performing method.
This highlights the advantage of MPC in combining partial tests across different sparsity levels, thereby increasing adaptability to shift patterns.
Moreover, as $p$ increases, the performance gap between MPC and competitors widens, particularly in Scenario II and Scenario III. This demonstrates the effectiveness of MPC, particularly in high-dimensional settings, in identifying potential sparse shifts characterized by weakened relationships between profiles.
It is important to note that Scenario II and Scenario III are more complex to detect than Scenario I and Scenario IV, particularly when $n_{el} = 1$. Indeed, although the severity level reaches its maximum when an edge is removed or the relationships are strongly weakened, the resulting precision matrix is less extreme than in the other two scenarios, where it approaches singularity.
When the process is IC, all methods exhibit $ARL$ values at least as large as the prespecified $ARL_0$ (100), demonstrating their ability to control false alarms under IC conditions. 
However, in these cases as well, MPC shows a more consistent decrease in $ARL$ with increasing shift severity and maintains acceptable detection power across a wide range of shift magnitudes.
The observed performance differences emphasize the importance of leveraging shift sparsity to capture the underlying structure of multichannel profiles.
The simulation results for Model II and Model III are similar to those presented earlier and can be found in the Supplementary Materials.

\section{Case Study}
\label{sec_real}

\begin{figure}
    \centering
    \includegraphics[width=1\linewidth]{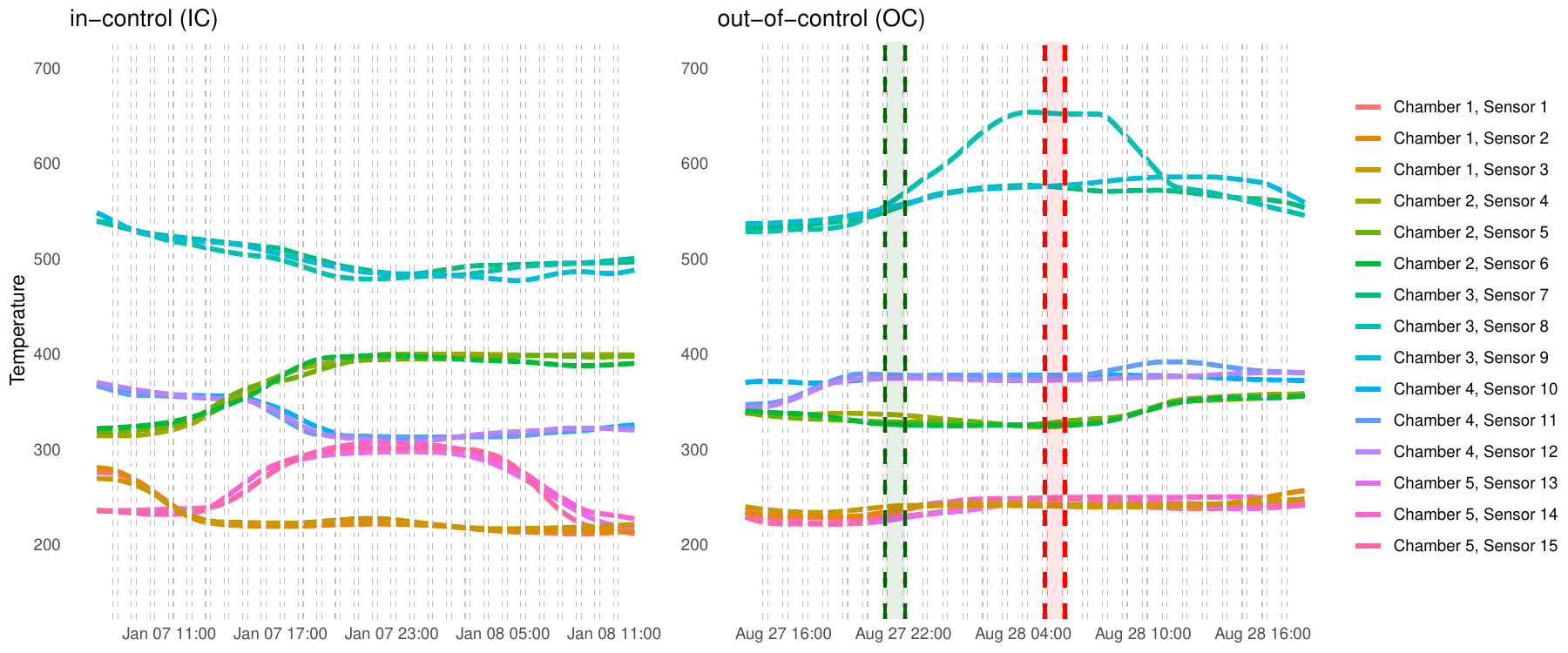}
    \caption{A sample of the original IC temperature data recorded from 2016-02-06 18:05:00 to 2016-02-07 22:05:00 (left panel). Original OC temperature data recorded from 2017-08-27 09:05:00 to 2017-08-28 13:05:00 (right panel), monitored in Phase II of the case study. The grey dashed vertical lines separate the hourly profiles for each product, while the red and green vertical bands highlight the stopping time and the change point identified by the proposed method, respectively.}
    \label{fig:icoc curves}
\end{figure}

The practical applicability of the proposed MPC control chart is demonstrated through a case study on the monitoring of a roasting process. Data are publicly available on Kaggle at \url{https://www.kaggle.com/datasets/podsyp/production-quality}.
The roasting machine consists of five chambers, each containing three sensors placed at distinct locations to obtain temperature profiles. As a result, $p=15$ temperature profiles can be collected for each roasting cycle, which lasts one hour. 
The data are collected at a rate of one temperature measurement per minute, yielding 60 data points per profile.
The dynamics of the temperature profiles are crucial to the quality of the final product, as measured by an overall score in the dataset that assesses industry standards for product size and hardness.
\begin{figure}
    \centering
    \includegraphics[width=1\linewidth]{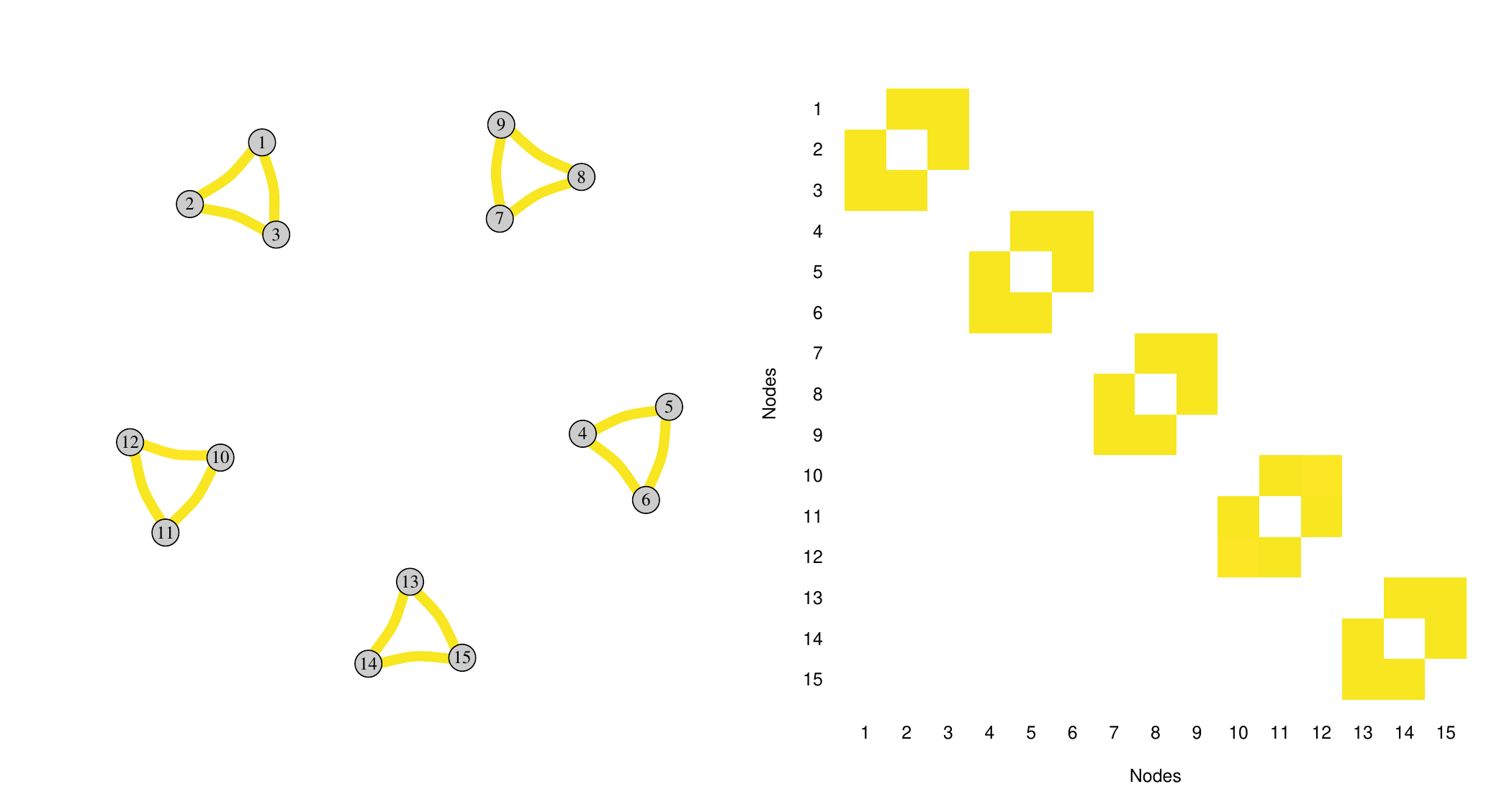}
    \caption{Estimated IC graph (left) and norms of the blocks of the estimated IC precision matrix (right) in the case study.}
    \label{fig:ic graph}
\end{figure}
Products with quality scores below the 20th percentile are classified as defective (5752 products). Accordingly, the remaining 21212 products are considered non-defective.
This dataset was also analyzed by \cite{yao2023adaptive}, who observed that the profiles associated with lower product-quality scores exhibit a shifted mean function compared to those associated with higher quality. This result was attained using a multi-profile control chart with an adaptive sampling strategy.
However, some sequences of OC profiles not only display a shift in the mean function but also reveal changes in the dependence structure among the temperature profiles. As discussed in Section \ref{sec_intro}, temperature profiles from sensors located within the same chamber are expected to behave similarly under IC conditions, as they capture the same physical phenomenon in a shared environment and thus can be considered multichannel data.
An illustrative example is shown in Figure \ref{fig:icoc curves}, where the left panel presents a sequence of 29 IC hourly profiles for each product, separated by grey dashed vertical lines, from the 15 sensors (channels) recorded from 2016-02-06 18:05:00 to 2016-02-07 22:05:00; analogously, the right panel reports a sequence of 28 OC hourly profiles collected from 2017-08-27 09:05:00 to 2017-08-28 13:05:00.
Consistent with the observations reported in the supplementary material of \cite{yao2023adaptive}, a clear mean shift in the sensors of the third chamber is visible, as shown in this figure. However, a noticeable change is evident in the relationships among sensor 8 and sensors 7 and 9, all located in the third chamber of the roasting machine.

Since the focus is on detecting potential shifts in process covariance, the data are preprocessed by removing the mean function calculated for each chamber.
Then, the resulting profiles for non-defective units were split equally into training and tuning sets and used in Phase I. As detailed in Algorithm \ref{alg:MPC}, the former set is used to estimate the MFPCA model and the IC precision matrix, whereas the latter is used to estimate the remaining parameters for Phase II. According to Section \ref{sec_sim}, the control limit $h$ is chosen such that $ARL_0 = 100$, the MEWMC parameter $\rho$ is set equal to 0.1, and the cardinality of $S$ is chosen as $n_s = 10$.
Figure \ref{fig:ic graph} shows the estimated IC precision matrix through the corresponding graph (left panel) and the $p\times p$ matrix of the norms of its elements (right panel). Sensors placed in each of the five chambers are clearly distinguishable, as their corresponding temperature profiles exhibit strong correlations. It is worth noting that profiles across chambers are conditionally independent, whereas profiles within the same chamber are not.

\begin{table}[]
\centering
\resizebox{0.55\linewidth}{!}{
\begin{tabular}{p{3cm}p{0.5cm}p{3cm}c}
\toprule
      &  & $\widehat{ARL}$ &  CI  \\ 
      \cline{3-4} \\[-0.7em]
MPC &  & \textbf{12.340}       & {[}10.611 , 14.068{]} \\  \\[-0.7em]
HGM  &     & 34.585       & {[}30.156 , 39.013{]} \\  \\[-0.7em]
Ren   &    & 17.710       & {[}15.439 , 19.980{]} \\ 
\bottomrule
\end{tabular}}
\caption{Estimated ARL values, denoted as $\widehat{ARL}$, on the Phase II sequences and the corresponding 95\% confidence interval (CI) for each monitoring method in the case study.}
\label{tab:arl case study}
\end{table}

To compare the performance of the MPC with that of the competing methods presented in Section \ref{sec_sim}, 200 OC sequences were randomly sampled from the 5752 OC profiles and monitored using each method. $ARL$ is calculated as the mean of the run lengths for each sequence.
The results are reported in Table \ref{tab:arl case study}, where the proposed MPC method appears to clearly outperform all competing methods and confirms the superior performance achieved in the simulation study.

\begin{figure}
    \centering
    \includegraphics[width=1\linewidth]{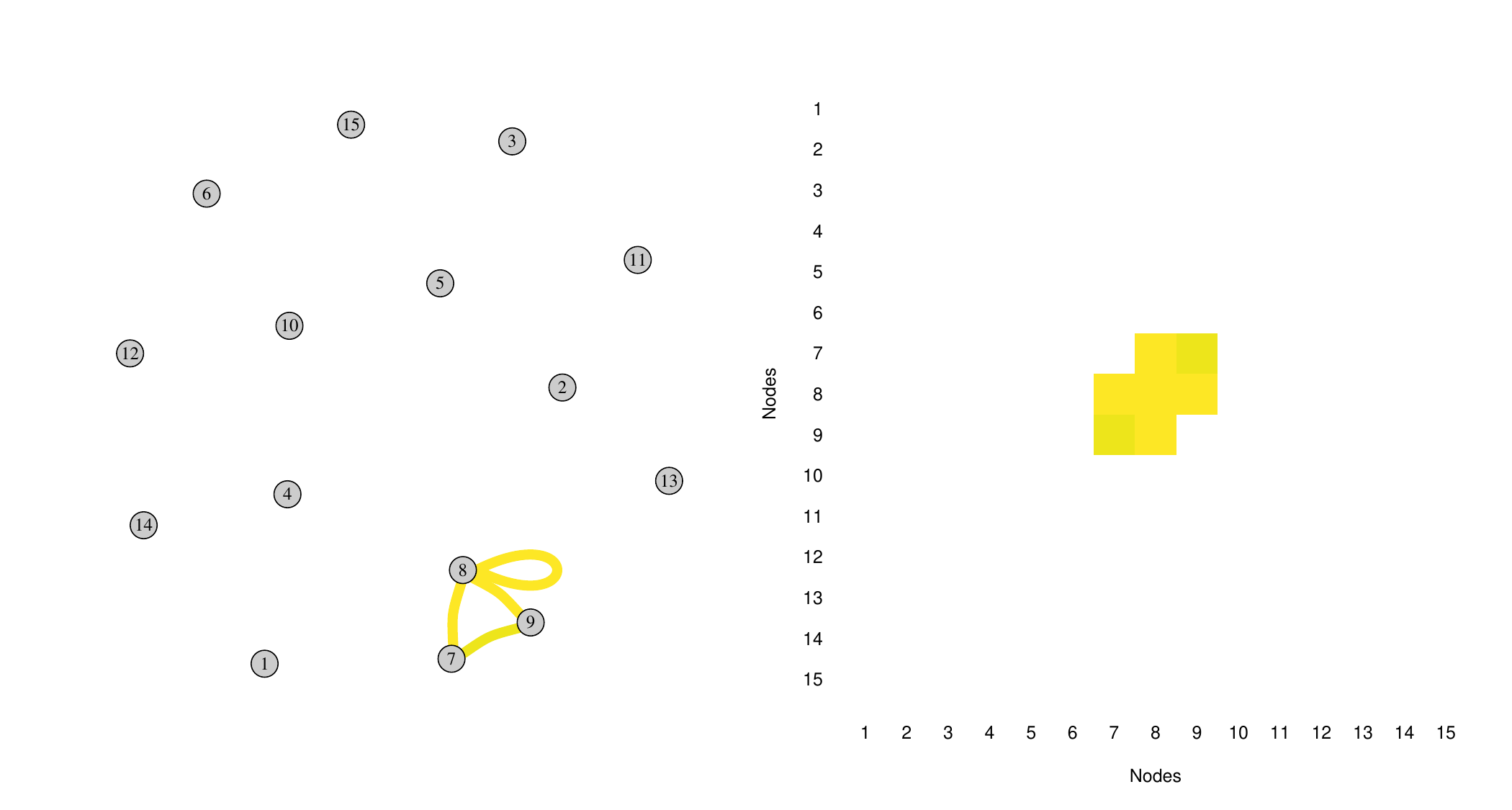}
    \caption{Graph of the significantly shifted conditional relationships (left) and matrix of the significant p-values (right) obtained with the BH procedure in the case study.}
    \label{fig:diffgraph}
\end{figure}

Finally, to demonstrate the effectiveness of the proposed method's post-signal diagnosis, we applied the proposed MPC control chart to the 28 OC profiles presented in Figure \ref{fig:icoc curves}, which detected a shift at $m = 65$, as highlighted by the red vertical band.
The post-signal diagnostic procedure detailed in Section \ref{post-signal} was then applied to all observations up to $m$, and it estimates a change point at $\widehat{\tau} = 57$, highlighted by the green vertical band.
Using the BH procedure described in Section \ref{post-signal} with $FDR = 0.01$, the shifted elements of the OC precision matrix are identified and represented in Figure \ref{fig:diffgraph}, with the left panel illustrating the associated graph.
The relationships involving sensor 8 and sensors 7 and 9 are correctly identified, suggesting a potential fault in that chamber and warranting further investigation by domain experts.
As demonstrated in this case study, the proposed MPC is a superior and interpretable tool for monitoring the covariance of multichannel profiles and identifying potential locations of shifts, thereby assisting domain experts in diagnosing the root causes of signalled process anomalies.

\section{Conclusions}
\label{sec_conclusions}
This article introduces the multichannel profile covariance (MPC) control chart for online monitoring of the covariance in multichannel profiles. The proposed method, leveraging the interpretable framework of functional graphical models, first encodes the conditional relationships among profiles in the in-control precision matrix, then performs online monitoring by identifying shifted blocks in the precision matrix and comparing multiple constrained likelihood ratio tests across different sparsity levels. In particular, the $s$ most significant shifted relationships are first identified using the corresponding $p$-values; the overall test statistic is then obtained by combining multiple tests across different sparsity levels using a nonparametric combination methodology. 
Through extensive Monte Carlo simulations, in which the proposed MPC control chart outperformed state-of-the-art methods, this combination is shown to be effective across a wide range of scenarios, enabling adaptation to unknown shift patterns.
Moreover, once the proposed control chart triggers a signal, the shift localization embedded in the monitoring procedure provides an immediate post-signal diagnostic that requires no additional computation, thereby informing practitioners which specific profile relationships may have shifted.
The practical applicability of the proposed method is illustrated through a case study on the monitoring of a roasting process. In this application, MPC not only detected an out-of-control state of the process compared with competing methods but also accurately identified the change point and provided interpretable diagnostics of the underlying shift, specifically isolating problematic relationships among sensors in the third chamber.

Several promising directions for future research emerge from this work. Although the current methodology relies on the assumption of multivariate Gaussian processes to establish the connection between the functional graphical model and the precision matrix of the principal component scores, extending the framework to nonparametric settings would significantly broaden its applicability. Moreover, the proposed method assumes that the observations are independent over time, which may not hold in many practical scenarios where temporal correlations exist between successive multichannel profiles. Incorporating temporal dependencies into the monitoring framework represents an important extension that could improve detection performance when such correlations are present. 
Finally, the current framework assumes that the process operates in a stationary regime under in-control conditions, with shifts occurring at unknown change points. However, many modern industrial processes exhibit nonstationary behavior due to gradual wear or slowly varying operating conditions. Developing a monitoring procedure that can distinguish between natural nonstationarity and anomalous shifts in the covariance structure would enhance the method's applicability in such contexts.

\section*{Supplementary Materials}
Supplementary Materials provide additional implementation details, simulation results, and R code to reproduce the results presented in the case study.

\bibliographystyle{apalike}
\setlength{\bibsep}{5pt plus 0.2ex}
{\small
\spacingset{1}
\bibliography{ref}

@article{ren2019phase,
  title={{Phase-II monitoring in multichannel profile observations}},
  author={Ren, Haojie and Chen, Nan and Wang, Zhaojun},
  journal={Journal of Quality Technology},
  volume={51},
  number={4},
  pages={338--352},
  year={2019},
  publisher={Taylor \& Francis}
}

@article{wu2022monitoring,
  title={Monitoring heterogeneous multivariate profiles based on heterogeneous graphical model},
  author={Wu, Hui and Zhang, Chen and Li, Yan-Fu},
  journal={Technometrics},
  volume={64},
  number={2},
  pages={210--223},
  year={2022},
  publisher={Taylor \& Francis}
}

@article{centofanti2025adaptive,
  title={An Adaptive Multivariate Functional Control Chart},
  author={Centofanti, Fabio and Lepore, Antonio and Palumbo, Biagio},
  journal={Technometrics},
  year={2025},
  publisher={Taylor \& Francis}
}

@article{li2018pairwise,
  title={Pairwise estimation of multivariate gaussian process models with replicated observations: Application to multivariate profile monitoring},
  author={Li, Yongxiang and Zhou, Qiang and Huang, Xiaohu and Zeng, Li},
  journal={Technometrics},
  volume={60},
  number={1},
  pages={70--78},
  year={2018},
  publisher={Taylor \& Francis}
}

@article{qiao2019functional,
  title={Functional graphical models},
  author={Qiao, Xinghao and Guo, Shaojun and James, Gareth M},
  journal={Journal of the American Statistical Association},
  volume={114},
  number={525},
  pages={211--222},
  year={2019},
  publisher={Taylor \& Francis}
}

@article{zapata2022partial,
  title={Partial separability and functional graphical models for multivariate Gaussian processes},
  author={Zapata, Javier and Oh, Sang-Yun and Petersen, Alexander},
  journal={Biometrika},
  volume={109},
  number={3},
  pages={665--681},
  year={2022},
  publisher={Oxford University Press}
}

@article{zhao2024high,
  title={High-dimensional functional graphical model structure learning via neighborhood selection approach},
  author={Zhao, Boxin and Zhai, Percy S. and Wang, Y. Samuel and Kolar, Mladen},
  journal={Electronic Journal of Statistics},
  volume={18},
  number={1},
  pages={143–188},
  year={2024}
}

@article{zhu2016bayesian,
  title={Bayesian graphical models for multivariate functional data},
  author={Zhu, Hongxiao and Strawn, Nate and Dunson, David B},
  journal={Journal of Machine Learning Research},
  volume={17},
  number={204},
  pages={1--27},
  year={2016}
}

@article{meinshausen2006high,
  title={High-dimensional graphs and variable selection with the Lasso},
  author={Meinshausen, Nicolai and B{\"u}hlmann, Peter},
  journal={Annals of Statistics},
  volume={34},
  number={3},
  pages={1436--1462},
  year={2006},
  publisher={Institute of Mathematical Statistics}
}

@article{friedman2008sparse,
  title={Sparse inverse covariance estimation with the graphical lasso},
  author={Friedman, Jerome and Hastie, Trevor and Tibshirani, Robert},
  journal={Biostatistics},
  volume={9},
  number={3},
  pages={432--441},
  year={2008},
  publisher={Oxford University Press}
}

@article{hawkins2008multivariate,
  title={Multivariate exponentially weighted moving covariance matrix},
  author={Hawkins, Douglas M and Maboudou-Tchao, Edgard M},
  journal={Technometrics},
  volume={50},
  number={2},
  pages={155--166},
  year={2008},
  publisher={Taylor \& Francis}
}

@article{maboudou2013lasso,
  title={A lasso chart for monitoring the covariance matrix},
  author={Maboudou-Tchao, Edgard M and Diawara, Norou},
  journal={Quality Technology \& Quantitative Management},
  volume={10},
  number={1},
  pages={95--114},
  year={2013},
  publisher={Taylor \& Francis}
}

@article{cai2011adaptive,
  title={Adaptive thresholding for sparse covariance matrix estimation},
  author={Cai, Tony and Liu, Weidong},
  journal={Journal of the American Statistical Association},
  volume={106},
  number={494},
  pages={672--684},
  year={2011},
  publisher={Taylor \& Francis}
}

@article{li2013monitoring,
  title={Monitoring the covariance matrix via penalized likelihood estimation},
  author={Li, Bo and Wang, Kaibo and Yeh, Arthur B},
  journal={{IIE Transactions}},
  volume={45},
  number={2},
  pages={132--146},
  year={2013},
  publisher={Taylor \& Francis}
}

@article{shen2014new,
  title={A new multivariate EWMA scheme for monitoring covariance matrices},
  author={Shen, Xiaobei and Tsung, Fugee and Zou, Changliang},
  journal={International Journal of Production Research},
  volume={52},
  number={10},
  pages={2834--2850},
  year={2014},
  publisher={Taylor \& Francis}
}

@article{yeh2012monitoring,
  title={Monitoring multivariate process variability with individual observations via penalised likelihood estimation},
  author={Yeh, Arthur B and Li, Bo and Wang, Kaibo},
  journal={International Journal of Production Research},
  volume={50},
  number={22},
  pages={6624--6638},
  year={2012},
  publisher={Taylor \& Francis}
}

@article{abdella2019adaptive,
  title={An adaptive thresholding-based process variability monitoring},
  author={Abdella, Galal M and Kim, Jinho and Kim, Sangahn and Al-Khalifa, Khalifa N and Jeong, Myong K and Hamouda, Abdel Magid and Elsayed, Elsayed A},
  journal={Journal of Quality Technology},
  volume={51},
  number={3},
  pages={242--256},
  year={2019},
  publisher={Taylor \& Francis}
}

@article{fan2009network,
title = {Network Exploration via the Adaptive LASSO and SCAD Penalties}, 
author = {Jianqing Fan and Yang Feng and Yichao Wu},
 journal = {The Annals of Applied Statistics},
 number = {2},
 pages = {521--541},
 publisher = {Institute of Mathematical Statistics},
 urldate = {2024-10-24},
 volume = {3},
 year = {2009}
}

@book{pesarin2010permutation,
  title={Permutation tests for complex data: theory, applications and software},
  author={Pesarin, Fortunato and Salmaso, Luigi},
  year={2010},
  publisher={John Wiley \& Sons}
}

@article{pesarin2010finite,
  title={Finite-sample consistency of combination-based permutation tests with application to repeated measures designs},
  author={Pesarin, Fortunato and Salmaso, Luigi},
  journal={Journal of Nonparametric Statistics},
  volume={22},
  number={5},
  pages={669--684},
  year={2010},
  publisher={Taylor \& Francis}
}

@article{jankova2015confidence,
  title={Confidence intervals for high-dimensional inverse covariance estimation},
  author={Jankova, Jana and van de Geer, Sara},
  journal={Electronic Journal of Statistics},
  volume={9},
  number={1},
  pages={1205--1229},
  year={2015},
  publisher={Institute of Mathematical Statistics}
}

@article{grasso2014profile,
  title={Profile monitoring via sensor fusion: the use of PCA methods for multi-channel data},
  author={Grasso, Marco and Colosimo, Bianca Maria and Pacella, Massimo},
  journal={International Journal of Production Research},
  volume={52},
  number={20},
  pages={6110--6135},
  year={2014},
  publisher={Taylor \& Francis}
}

@article{wang2018thresholded,
  title={Thresholded multivariate principal component analysis for phase I multichannel profile monitoring},
  author={Wang, Yuan and Mei, Yajun and Paynabar, Kamran},
  journal={Technometrics},
  volume={60},
  number={3},
  pages={360--372},
  year={2018},
  publisher={Taylor \& Francis}
}

@article{paynabar2016change,
  title={{A change-point approach for phase-I analysis in multivariate profile monitoring and diagnosis}},
  author={Paynabar, Kamran and Zou, Changliang and Qiu, Peihua},
  journal={Technometrics},
  volume={58},
  number={2},
  pages={191--204},
  year={2016},
  publisher={Taylor \& Francis}
}

@article{zhang2018weakly,
  title={Weakly correlated profile monitoring based on sparse multi-channel functional principal component analysis},
  author={Zhang, Chen and Yan, Hao and Lee, Seungho and Shi, Jianjun},
  journal={IISE Transactions},
  volume={50},
  number={10},
  pages={878--891},
  year={2018},
  publisher={Taylor \& Francis}
}

@article{paynabar2013monitoring,
  title={Monitoring and diagnosis of multichannel nonlinear profile variations using uncorrelated multilinear principal component analysis},
  author={Paynabar, Kamran and Jin, Jionghua and Pacella, Massimo},
  journal={IIE Transactions},
  volume={45},
  number={11},
  pages={1235--1247},
  year={2013},
  publisher={Taylor \& Francis}
}

@article{zhang2018multiple,
  title={Multiple profiles sensor-based monitoring and anomaly detection},
  author={Zhang, Chen and Yan, Hao and Lee, Seungho and Shi, Jianjun},
  journal={Journal of Quality Technology},
  volume={50},
  number={4},
  pages={344--362},
  year={2018},
  publisher={Taylor \& Francis}
}

@article{capezza2024robust,
  title={Robust multivariate functional control chart},
  author={Capezza, Christian and Centofanti, Fabio and Lepore, Antonio and Palumbo, Biagio},
  journal={Technometrics},
  year={2024},
  volume = {66},
  number = {4},
  pages = {531--547},
  publisher={Taylor \& Francis}
}

@book{qiu2013introduction,
  title={Introduction to statistical process control},
  author={Qiu, Peihua},
  year={2014},
  publisher={CRC press}
}

@book{ramsay,
  title={Functional data analysis},
  author={Ramsay, J.O. and Silverman, B.W.},
  year={2005},
  publisher={Springer}
}

@book{kokoszka,
  title={Introduction to functional data analysis},
  author={Kokoszka, Piotr and Reimherr, Matthew},
  year={2017},
  publisher={Chapman and Hall/CRC}
}

@book{noorossana2011statistical,
  title={Statistical analysis of profile monitoring},
  author={Noorossana, Rassoul and Saghaei, Abbas and Amiri, Amirhossein},
  year={2011},
  publisher={John Wiley \& Sons}
}

@article{horvath2014testing,
  title={Testing stationarity of functional time series},
  author={Horv{\'a}th, Lajos and Kokoszka, Piotr and Rice, Gregory},
  journal={Journal of Econometrics},
  volume={179},
  number={1},
  pages={66--82},
  year={2014},
  publisher={Elsevier}
}

@article{keshavarz2020sequential,
  title={Sequential change-point detection in high-dimensional Gaussian graphical models},
  author={Keshavarz, Hossein and Michaildiis, George and Atchad{\'e}, Yves},
  journal={Journal of machine learning research},
  volume={21},
  number={82},
  pages={1--57},
  year={2020}
}

@article{avanesov2018change,
  title={Change-point detection in high-dimensional covariance structure},
  author={Avanesov, Valeriy and Buzun, Nazar},
  journal={Electronic Journal of Statistics},
  volume={12},
  number={2},
  pages={3254--3294},
  year={2018},
  publisher={Institute of Mathematical Statistics}
}

@article{zhao2020distributed,
  title={Distributed acoustic sensing vertical seismic profile data denoiser based on convolutional neural network},
  author={Zhao, Yuxing and Li, Yue and Wu, Ning},
  journal={IEEE Transactions on Geoscience and Remote Sensing},
  volume={60},
  pages={1--11},
  year={2020},
  publisher={IEEE}
}

@article{jensen2006effects,
  title={Effects of parameter estimation on control chart properties: a literature review},
  author={Jensen, Willis A and Jones-Farmer, L Allison and Champ, Charles W and Woodall, William H},
  journal={Journal of Quality technology},
  volume={38},
  number={4},
  pages={349--364},
  year={2006},
  publisher={Taylor \& Francis}
}

@Manual{r2021,
title = {R: A Language and Environment for Statistical Computing},
author = {{R Core Team}},
organization = {R Foundation for Statistical Computing},
address = {Vienna, Austria},
year = {2024},
url = {https://www.R-project.org/},
}

@book{kruger2012statistical,
title={Statistical monitoring of complex multivatiate processes: with applications in industrial process control},
author={Kruger, Uwe and Xie, Lei},
year={2012},
publisher={John Wiley \& Sons}
}

@article{yao2023adaptive,
  title={Adaptive sampling for monitoring multi-profile data with within-and-between profile correlation},
  author={Yao, Jinwei and Xian, Xiaochen and Wang, Chao},
  journal={Technometrics},
  volume={65},
  number={3},
  pages={375--387},
  year={2023},
  publisher={Taylor \& Francis}
}

@article{boyd2011distributed,
  title={Distributed optimization and statistical learning via the alternating direction method of multipliers},
  author={Boyd, Stephen and Parikh, Neal and Chu, Eric and Peleato, Borja and Eckstein, Jonathan and others},
  journal={Foundations and Trends{\textregistered} in Machine learning},
  volume={3},
  number={1},
  pages={1--122},
  year={2011},
  publisher={Now Publishers, Inc.}
}

@article{qiu2003nonparametric,
  title={A nonparametric multivariate cumulative sum procedure for detecting shifts in all directions},
  author={Qiu, Peihua and Hawkins, Douglas},
  journal={Journal of the Royal Statistical Society Series D: The Statistician},
  volume={52},
  number={2},
  pages={151--164},
  year={2003},
  publisher={Oxford University Press}
}

@article{lim2024efficient,
  title={Efficient ARL estimation for general control charts using censored run lengths},
  author={Lim, Johan and Lee, Sungim},
  journal={Quality Engineering},
  pages={1--10},
  year={2024},
  publisher={Taylor \& Francis}
}

@article{van2016ridge,
  title={Ridge estimation of inverse covariance matrices from high-dimensional data},
  author={Van Wieringen, Wessel N and Peeters, Carel FW},
  journal={Computational Statistics \& Data Analysis},
  volume={103},
  pages={284--303},
  year={2016},
  publisher={Elsevier}
}

@incollection{fisher1970statistical,
  title={Statistical methods for research workers},
  author={Fisher, Ronald Aylmer},
  booktitle={Breakthroughs in statistics: Methodology and distribution},
  pages={66--70},
  year={1970},
  publisher={Springer}
}

@article{capezza2025adaptive,
  title={An adaptive multivariate functional EWMA control chart},
  author={Capezza, Christian and Capizzi, Giovanna and Centofanti, Fabio and Lepore, Antonio and Palumbo, Biagio},
  journal={Journal of Quality Technology},
  volume={57},
  number={1},
  pages={1--15},
  year={2025},
  publisher={Taylor \& Francis}
}

@article{benjamini1995controlling,
  title={Controlling the false discovery rate: a practical and powerful approach to multiple testing},
  author={Benjamini, Yoav and Hochberg, Yosef},
  journal={Journal of the Royal statistical society: series B (Methodological)},
  volume={57},
  number={1},
  pages={289--300},
  year={1995},
  publisher={Wiley Online Library}
}

@article{zhao2022fudge,
  title={FuDGE: A method to estimate a functional differential graph in a high-dimensional setting},
  author={Zhao, Boxin and Wang, Y Samuel and Kolar, Mladen},
  journal={Journal of Machine Learning Research},
  volume={23},
  number={82},
  pages={1--82},
  year={2022}
}
}
\end{document}


\def\spacingset#1{\renewcommand{\baselinestretch}%
{#1}\small\normalsize} \spacingset{1}
\newcommand{\specialcell}[2][c]{%
  \begin{tabular}[#1]{@{}c@{}}#2\end{tabular}}


\if0\blind
{
  
\title{Supplementary Materials to ``Monitoring Covariance in Multichannel Profiles via Functional Graphical Models''}


\author[]{Christian Capezza}
\author[]{Davide Forcina}
\author[]{Antonio Lepore\thanks{Corresponding author. e-mail: \texttt{antonio.lepore@unina.it}}}
\author[]{Biagio Palumbo}

\affil[]{Department of Industrial Engineering, University of Naples Federico II, Piazzale Tecchio 80, 80125, Naples, Italy}

\setcounter{Maxaffil}{0}
\renewcommand\Affilfont{\itshape\small}
\date{}
\maketitle
} \fi

\if1\blind
{
  \bigskip
  \bigskip
  \bigskip
  \begin{center}
    {\LARGE\bf Supplementary Materials to ``Monitoring Covariance in Multichannel Profiles via Functional Graphical Models''}
\end{center}
  \medskip
} \fi

\vfill
\hfill {\tiny technometrics tex template (do not remove)}

\newpage
\spacingset{1.45} 
\appendix
\numberwithin{equation}{section}

\section{Additional Simulation Results}
\label{sec_appA}
\begin{figure}
    \centering
    \includegraphics[width=1\linewidth]{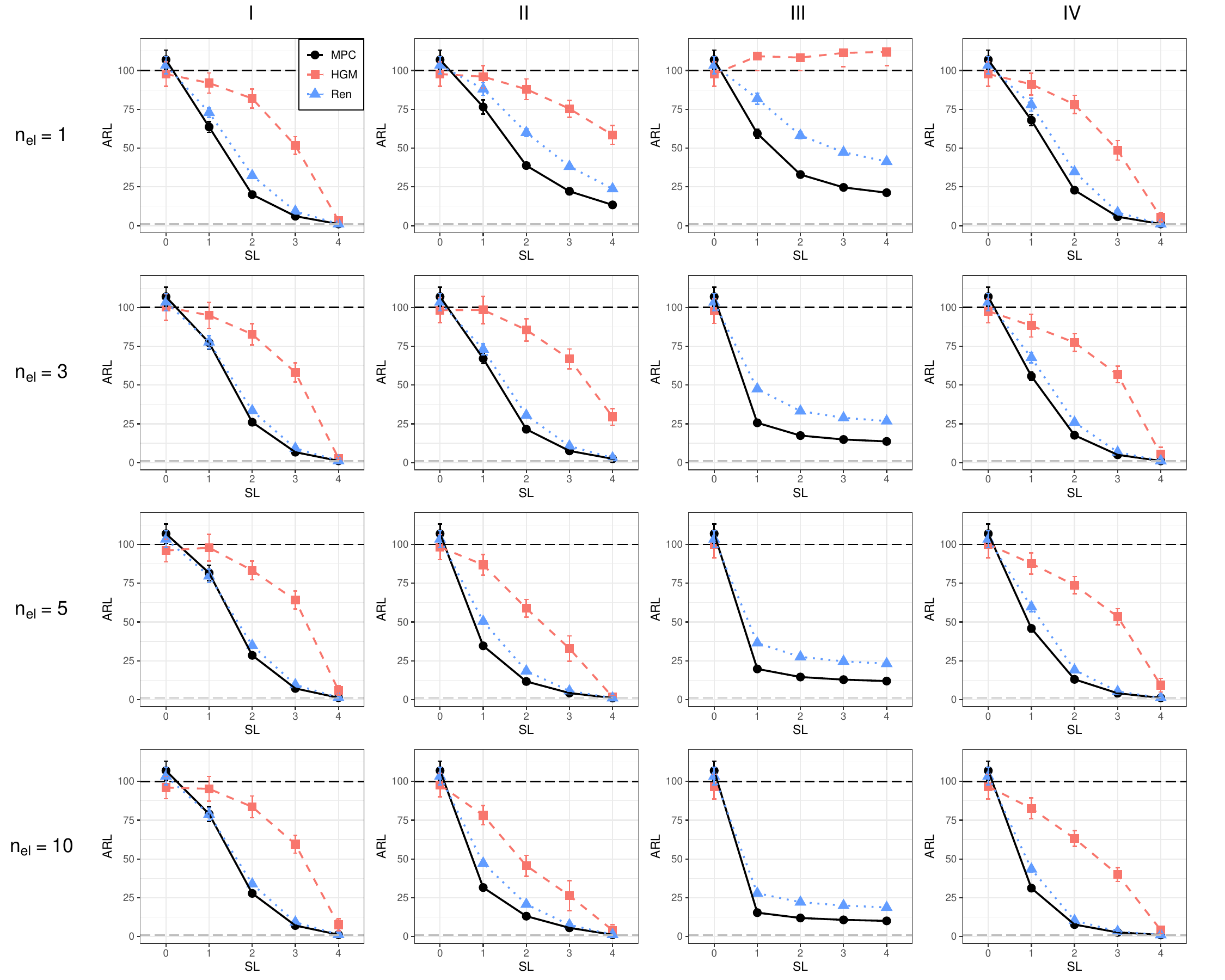}
    \caption{Mean $ARL$ achieved in Phase II by MPC, HGM, and Ren for Model I and $p=20$ functional variables as a function of SL.}
    \label{fig:p_20_graph_I}
\end{figure}

\begin{figure}
    \centering
    \includegraphics[width=1\linewidth]{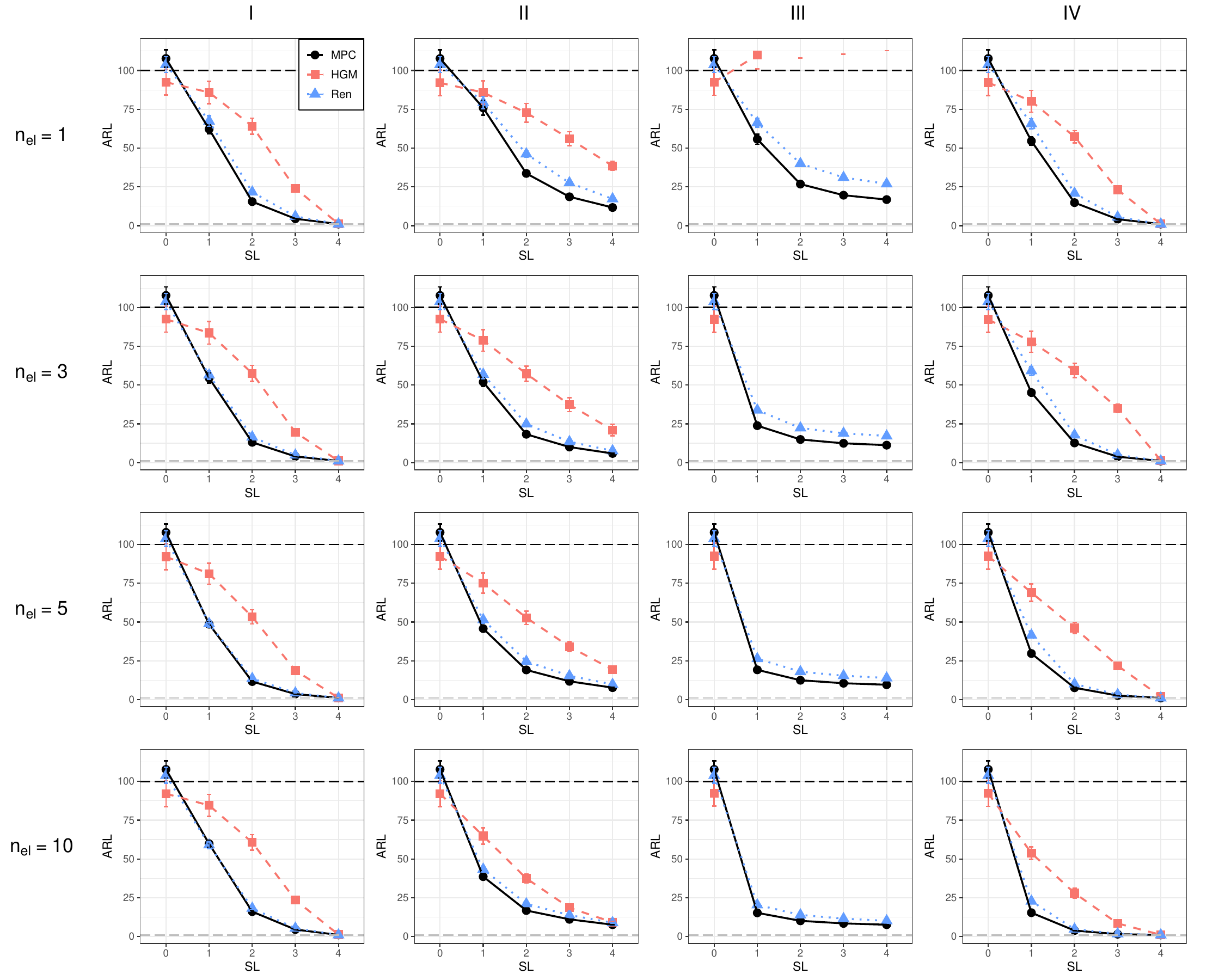}
    \caption{Mean $ARL$ achieved in Phase II by MPC, HGM, and Ren for Model II and $p=10$ functional variables as a function of SL.}
    \label{fig:p_10_graph_II}
\end{figure}

\begin{figure}
    \centering
    \includegraphics[width=1\linewidth]{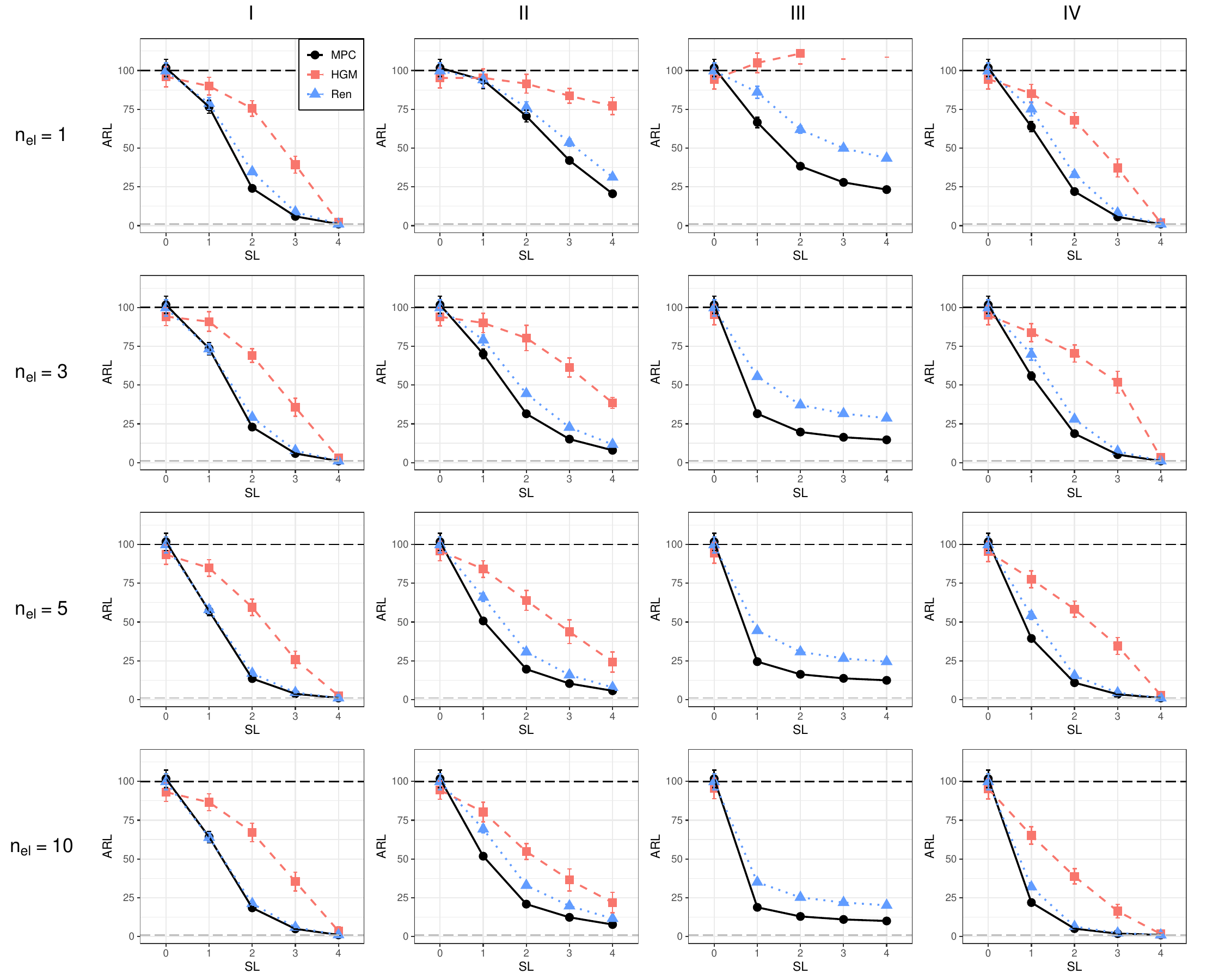}
    \caption{Mean $ARL$ achieved in Phase II by MPC, HGM, and Ren for Model II and $p=20$ functional variables as a function of SL.}
    \label{fig:p_20_graph_II}
\end{figure}

\begin{figure}
    \centering
    \includegraphics[width=1\linewidth]{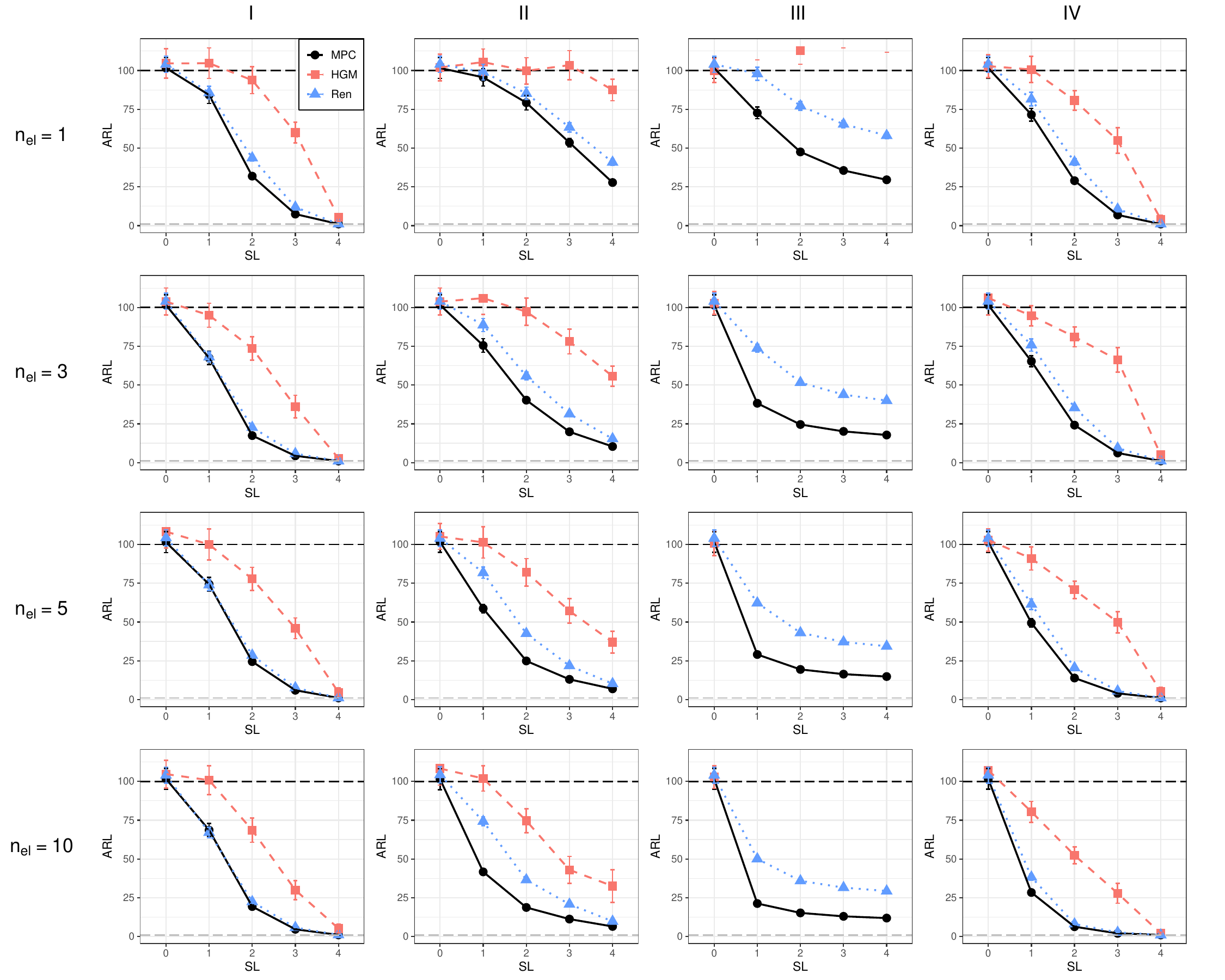}
    \caption{Mean $ARL$ achieved in Phase II by MPC, HGM, and Ren for Model II and $p=30$ functional variables as a function of SL.}
    \label{fig:p_30_graph_II}
\end{figure}

\begin{figure}
    \centering
    \includegraphics[width=1\linewidth]{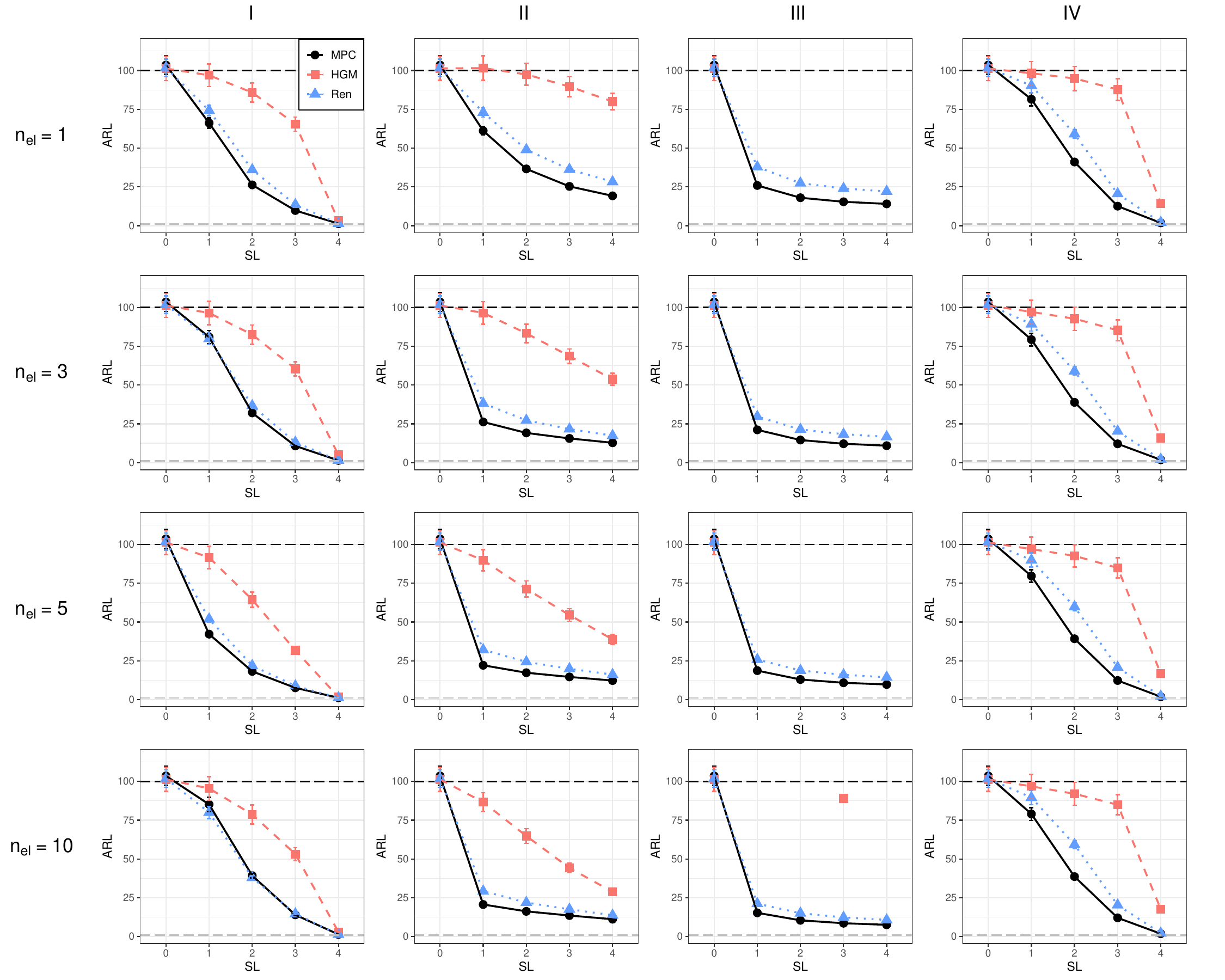}
    \caption{Mean $ARL$ achieved in Phase II by MPC, HGM, and Ren for Model III and $p=10$ functional variables as a function of SL.}
    \label{fig:p_10_graph_III}
\end{figure}

\begin{figure}
    \centering
    \includegraphics[width=1\linewidth]{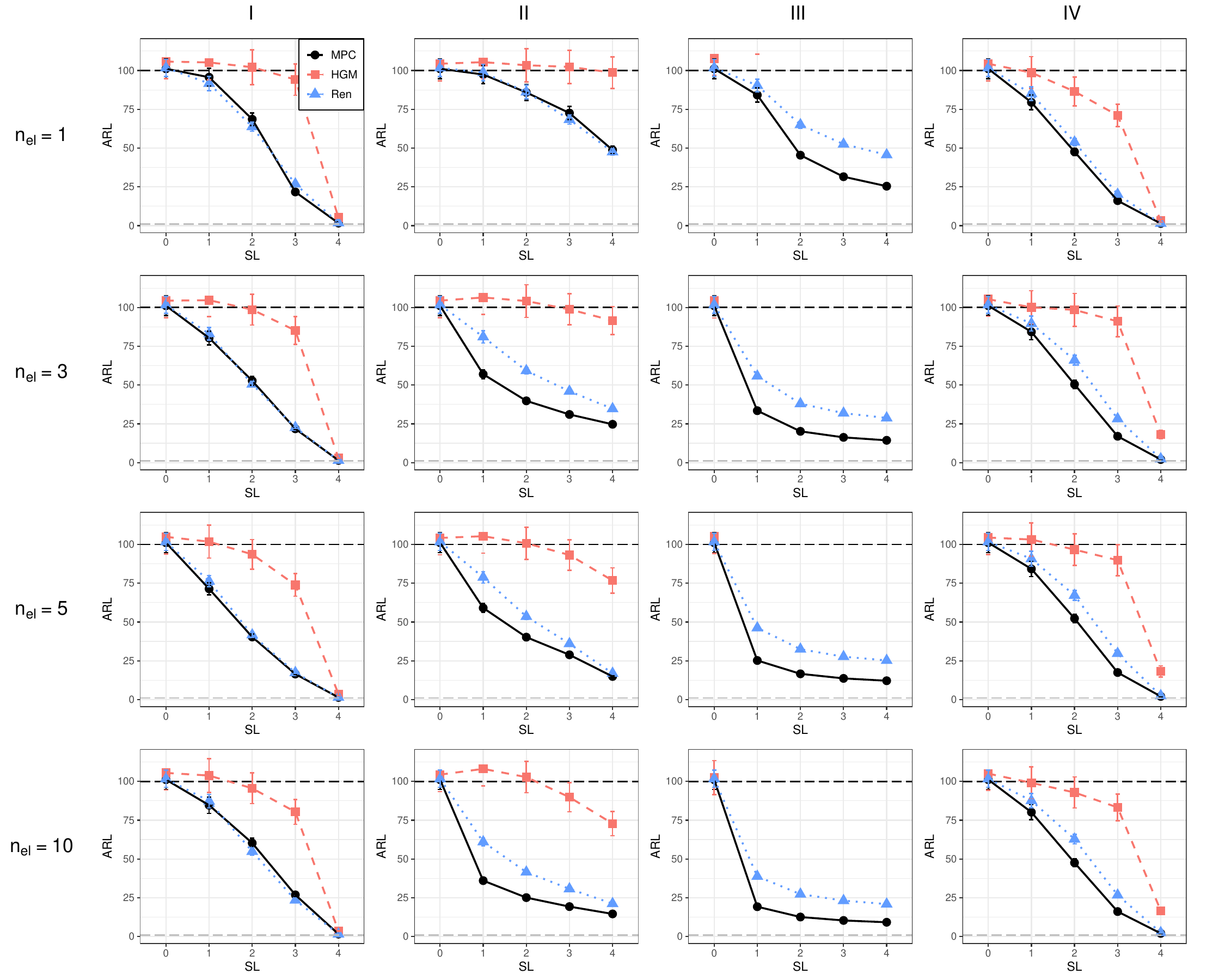}
    \caption{Mean $ARL$ achieved in Phase II by MPC, HGM, and Ren for Model III and $p=20$ functional variables as a function of SL.}
    \label{fig:p_20_graph_III}
\end{figure}

\begin{figure}
    \centering
    \includegraphics[width=1\linewidth]{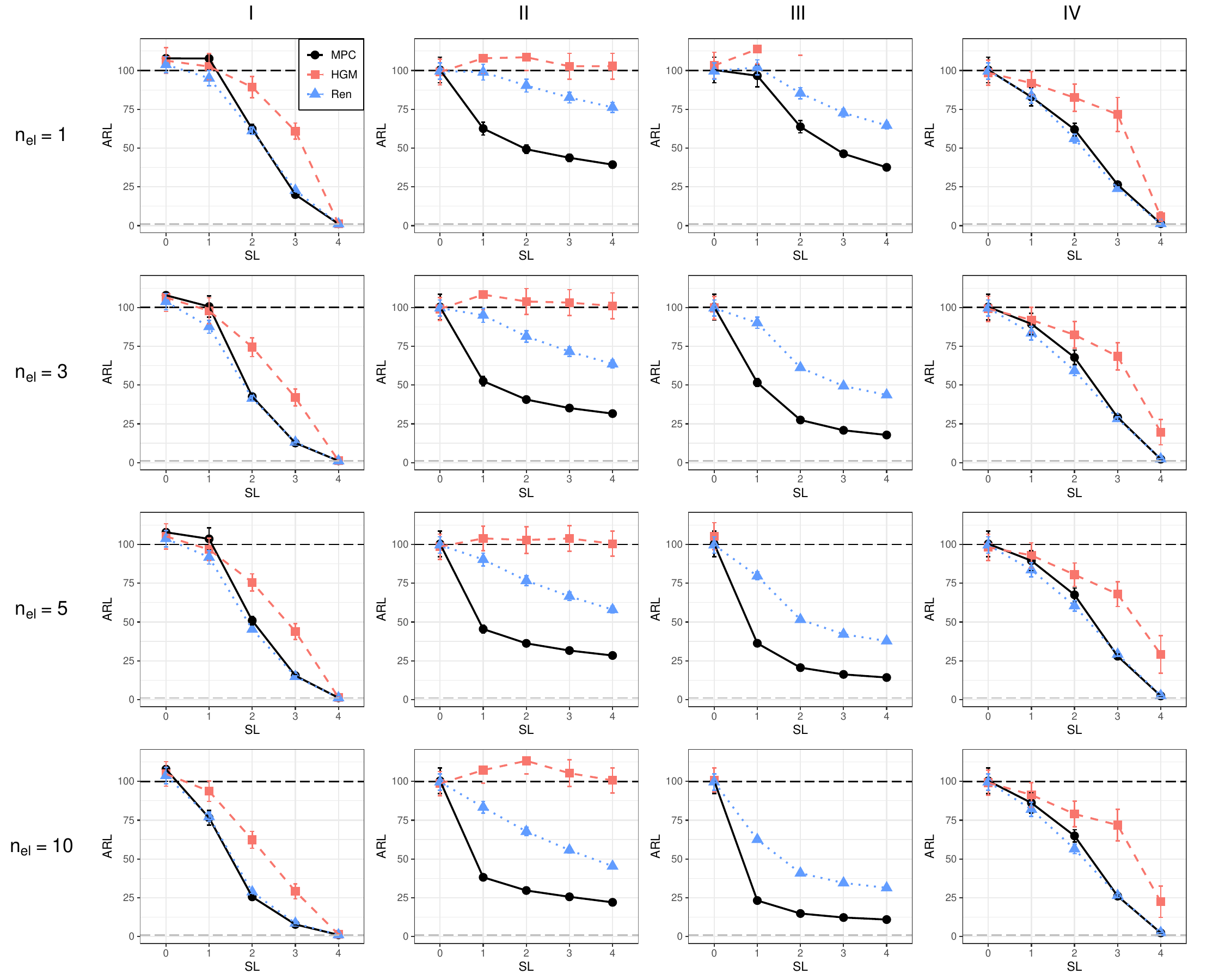}
    \caption{Mean $ARL$ achieved in Phase II by MPC, HGM, and Ren for Model III and $p=30$ functional variables as a function of SL.}
    \label{fig:p_30_graph_III}
\end{figure}

This Section provides simulation results for the $p=20$ functional variables and Model I settings, as well as simulation results for Model II and Model III presented in Section \ref{sec_sim}.
Figure \ref{fig:p_20_graph_I} shows the simulation results for the case of $p = 20$ functional variables and Model I, which are analogous to those presented in Section \ref{sec_sim} for the case of $p = 10$ and $p = 30$ functional variables. The proposed method outperforms all competing methods, with a larger performance gap in Scenario II and Scenario III. The performance gap between competing methods is observed only between cases with $p = 10$ and $p = 30$ functional variables, indicating that as $p$ increases, the proposed MPC control chart gains an increasing advantage over Ren and HGM.
The simulation results for Model II are shown in Figures \ref{fig:p_10_graph_II}, \ref{fig:p_20_graph_II}, and \ref{fig:p_30_graph_II}. As discussed in Section \ref{sec_sim}, the proposed MPC outperforms all competing methods. The performance gap becomes increasingly pronounced as the number of variables $p$ increases.
Finally, Figures \ref{fig:p_10_graph_III}, \ref{fig:p_20_graph_III}, and \ref{fig:p_30_graph_III} show simulation results for Model III. In this case, the proposed method continues to outperform all competing methods in Scenarios II, III, and IV, with a substantial gap relative to Model I and Model II when $p = 10$.
For $p = 20$ and $p = 30$, the performance gap relative to competing methods widens, except in Scenario I, where the performance of the proposed MPC and Ren is comparable. As already mentioned in Section \ref{sec_sim}, this scenario is less challenging with respect to Scenarios II and III, because the precision matrix starts to get close to singularity and, therefore, is more extreme. However, the performance gain of MPC over Ren in all other scenarios is substantially greater, which makes the proposed method the best choice for monitoring the covariance of multichannel profiles.


\section{ADMM algorithm}
\label{sec_appB}
To solve Equation \eqref{afglasso} and Equation \eqref{constrfgm}, an optimization algorithm should be used. In this article, the alternate direction method of multipliers (ADMM) \citep{boyd2011distributed} is adopted, as it guarantees the estimator of the precision matrix to be positive definite, otherwise not usable in the log-likelihood in Equation \eqref{loglk}. We first focus our attention on solving Equation \eqref{afglasso}, as the solution of Equation \eqref{constrfgm} can be easily obtained from it.
Following \cite{zhao2022fudge}, Equation \eqref{afglasso} can be reformulated as
\begin{equation*}
    \widehat{\boldsymbol{\Theta}}_{0} = \argmin_{\boldsymbol{\Theta}}\biggl\{-\log|\boldsymbol{\Theta}| + \Tr\left(\boldsymbol{P}\widehat{\boldsymbol{\Sigma}}_0\boldsymbol{P}^T\boldsymbol{\Theta} \right)  + \Gamma\left(\boldsymbol{Z}\right)\biggl\},
\end{equation*}
subject to $\boldsymbol{\Theta} \succ 0$ and $\boldsymbol{Z} = \boldsymbol{\Theta}$. The scaled augmented Lagrangian \citep{boyd2011distributed} is
\begin{equation}
\label{augmented lag}
    L_\rho \left(\{\boldsymbol{\Theta}\},\{\boldsymbol{Z}\},\{\boldsymbol{U}\}\right) = -\log|\boldsymbol{\Theta}| + \Tr\left(\boldsymbol{P}\widehat{\boldsymbol{\Sigma}}_0\boldsymbol{P}^T\boldsymbol{\Theta} \right)  + \Gamma\left(\boldsymbol{Z}\right) + \frac{\rho}{2}||\boldsymbol{\Theta} - \boldsymbol{Z} + \boldsymbol{U}||_F^2,
\end{equation}
where $\rho \geq 0$ is a tuning parameter and $\boldsymbol{U}$ is the dual variable.
The ADMM will solve Equation \eqref{augmented lag}, iterating the following steps until convergence. In iteration $i$:
\begin{enumerate}
    \item $\{\boldsymbol{\Theta}_{(i)}\} \leftarrow \argmin_{\{\boldsymbol{\Theta}\}} L_\rho \left(\{\boldsymbol{\Theta}\},\{\boldsymbol{Z}_{(i-1)}\},\{\boldsymbol{U}_{(i-1)}\}\right)$.
    \item $\{\boldsymbol{Z}_{(i)}\} \leftarrow \argmin_{\{\boldsymbol{Z}\}} L_\rho \left(\{\boldsymbol{\Theta}_{(i)}\},\{\boldsymbol{Z}\},\{\boldsymbol{U}_{(i-1)}\}\right)$.
    \item $\{\boldsymbol{U}_{(i)}\} \leftarrow \{\boldsymbol{U}_{(i-1)}\} + \{\boldsymbol{\Theta}_{(i)}\} - \{\boldsymbol{Z}_{(i)}\}$.
\end{enumerate}
The solution $\boldsymbol{\Theta}_{(i)}$ of step 1 is obtained as the minimizer of
\begin{equation}
\label{step1}
    -\log|\boldsymbol{\Theta}| + \Tr\left(\boldsymbol{P}\widehat{\boldsymbol{\Sigma}}_0\boldsymbol{P}^T\boldsymbol{\Theta} \right)  + \frac{\rho}{2}||\boldsymbol{\Theta} - \boldsymbol{Z}_{(i-1)} + \boldsymbol{U}_{(i-1)}||_F^2.
\end{equation}
Let $VDV^T$ be the eigendecomposition of $\boldsymbol{P}\widehat{\boldsymbol{\Sigma}}_0\boldsymbol{P}^T - \rho\boldsymbol{Z}_{(i-1)} + \rho\boldsymbol{U}_{(i-1)}$. The solution is given by $V\tilde{D}V^T$, where $\tilde{D}$ is the diagonal matrix with the $j$-th diagonal element being
\begin{equation*}
    \frac{1}{2\rho}\left(-D_{jj}+\sqrt{D_{jj}^2 + 4\rho} \right),
\end{equation*}
with $D_{jj}$ the $(j,j)$-th entry of $D$.
However, a more computationally convenient way to solve step 1 can be obtained by considering the permutation of the elements of $\boldsymbol{\Theta}$ induced by $\boldsymbol{P}^T$. In this case, the solution $\boldsymbol{\Theta}_{(i)}$ of step 1, is  obtained by minimizing
\begin{equation}
\label{step1transformed}
    -\log|\boldsymbol{P}^T\boldsymbol{\Theta}\boldsymbol{P}| + \Tr\left(\widehat{\boldsymbol{\Sigma}}_0\boldsymbol{P}^T\boldsymbol{\Theta}\boldsymbol{P} \right)  + \frac{\rho}{2}||\boldsymbol{P}^T(\boldsymbol{\Theta} - \boldsymbol{Z}_{(i-1)} + \boldsymbol{U}_{(i-1)})\boldsymbol{P}||_F^2,
\end{equation}
where the transformed matrices are block diagonal matrices with $K\times K$ blocks.
With this transformation, the solution to Equation \eqref{step1transformed} can be obtained, as shown above, for each block separately.
To be specific, let $V_kD_kV_k^T$ be the eigendecomposition of $\widehat{\boldsymbol{\Sigma}}_{0,k} - \rho (\boldsymbol{P}^T\boldsymbol{Z}_{(i-1)}\boldsymbol{P})_k + \rho (\boldsymbol{P}^T\boldsymbol{U}_{(i-1)}\boldsymbol{P})_k$, $k = 1,\dots,K$. The solution is given by $V_k\tilde{D}_kV^T_k$, where $\tilde{D}_k$ is the diagonal matrix with the $j$-th diagonal element being
\begin{equation*}
    \frac{1}{2\rho}\left(-D_{k,jj}+\sqrt{D_{k,jj}^2 + 4\rho} \right),
\end{equation*}
with $D_{k,jj}$ the $(j,j)$-th entry of $D_k$.
Then, $\boldsymbol{\Theta}_{(i)}$ is obtained as
\begin{equation*}
    \boldsymbol{\Theta}_{(i)} = \boldsymbol{P}V\tilde{D}V^T\boldsymbol{P}^T,
\end{equation*}
where $V = (V_1^T,\dots,V_K^T)^T$ and $\tilde{D}$ is the block-diagonal matrix with blocks $\tilde{D}_k$, $k = 1,\dots,K$. With the permutation induced by $\boldsymbol{P}^T$, the computational cost of step 1 is reduced from $O((Kp)^3)$ to $O(Kp^3)$. 

The solution $\boldsymbol{Z}_{(i)}$ of step 2 is obtained as the minimizer of
\begin{equation}
\label{step2}
     \frac{\rho}{2}||\boldsymbol{\Theta}_{(i)} - \boldsymbol{Z} + \boldsymbol{U}_{(i-1)}||_F^2 + \Gamma\left(\boldsymbol{Z}\right),
\end{equation}
where
\begin{equation*}
    \Gamma(\boldsymbol{Z}) = \lambda\sum_{j \leq l}\frac{||\boldsymbol{Z}_{jl}||_{\text{F}}}{||\tilde{\boldsymbol{\Theta}}_{0jl}||_{\text{F}}} = \lambda\sum_{j \leq l}\frac{||\boldsymbol{Z}_{jl}||_{\text{F}}}{w_{jl}}.
\end{equation*}
By defining $\boldsymbol{A} = \boldsymbol{\Theta}_{(i)} + \boldsymbol{U}_{(i-1)}$ and changing the sign of the term in the norm, we can rewrite Equation \eqref{step2} as
\begin{equation}
\label{step2sub}
     \frac{\rho}{2}||\boldsymbol{Z} - \boldsymbol{A}||_F^2 + \lambda\sum_{j \leq l}\frac{||\boldsymbol{Z}_{jl}||_{\text{F}}}{w_{jl}}.
\end{equation}
For any $1 \leq j,l,\leq p$, we can obtain the minimizer of Equation \eqref{step2sub}, denoted by $\widehat{\boldsymbol{Z}}_{j,l}$, by solving
\begin{equation}
\label{Zhatsol}
     \argmin_{\boldsymbol{Z}_{jl}}\left\{\frac{\rho}{2}||\boldsymbol{Z}_{jl} - \boldsymbol{A}_{jl}||_F^2 + \lambda\frac{||\boldsymbol{Z}_{jl}||_{\text{F}}}{w_{jl}}\right\}.
\end{equation}
Denoting the objective function of Equation \eqref{Zhatsol} by $L_{jl}$. Then, the subdifferential of $L_{jl}$ w.r.t. $\boldsymbol{Z}_{jl}$ is
\begin{equation*}
    \partial_{\boldsymbol{Z}_{jl}}L_{jl} = \rho(\boldsymbol{Z}_{jl} - \boldsymbol{A}_{jl}) + \frac{\lambda}{w_{jl}} \boldsymbol{G}_{jl},
\end{equation*}
where
\begin{equation*}
    \boldsymbol{G}_{jl} = \begin{cases}
    \frac{\boldsymbol{Z}_{jl}}{\|\boldsymbol{Z}_{jl}\|_F} & \text{if } \boldsymbol{Z}_{jl} \neq 0, \\[10pt]
    \left\{ \boldsymbol{G}_{jl} \in \mathbb{R}^{K \times K} : \|\boldsymbol{G}_{jl}\|_F \leq 1 \right\}, & \text{if }  \boldsymbol{Z}_{jl} = 0.
        \end{cases}
\end{equation*}
To obtain the optimum, we need
\begin{equation*}
    0 \in \partial_{\boldsymbol{Z}_{jl}}L_{jl}(\widehat{\boldsymbol{Z}}_{jl}).
\end{equation*}
\\
Suppose $\widehat{\boldsymbol{Z}}_{jl} = 0$. In this case, there exist $\boldsymbol{G}_{jl}$, with $\|\boldsymbol{G}_{jl}\|_F\leq 1$, such that
\begin{equation*}
    -\rho\boldsymbol{A}_{jl} + \frac{\lambda}{w_{jl}}\boldsymbol{G}_{jl} = 0.
\end{equation*}
This implies that
\begin{equation*}
    \boldsymbol{G}_{jl} = \frac{\rho w_{jl}}{\lambda}\boldsymbol{A}_{jl}
\end{equation*}
Therefore, we have
\begin{equation*}
    \|\boldsymbol{G}_{jl}\|_F = \frac{\rho w_{jl}}{\lambda}\|\boldsymbol{A}_{jl}\|_F\leq 1.
\end{equation*}
Thus
\begin{equation}
\label{claim1cond1}
    \|\boldsymbol{A}_{jl}\|_F\leq \frac{\lambda}{\rho w_{jl}}.
\end{equation}
Now, suppose $\widehat{\boldsymbol{Z}}_{jl} \neq 0$. In this case, we have
\begin{equation*}
    \rho\widehat{\boldsymbol{Z}}_{jl}-\rho\boldsymbol{A}_{jl} + \frac{\lambda}{w_{jl}}\frac{\widehat{\boldsymbol{Z}}_{jl}}{\|\widehat{\boldsymbol{Z}}_{jl}\|_F} = 0,
\end{equation*}
which implies
\begin{equation}
\label{claim2cond1}
    \boldsymbol{A}_{jl} = \widehat{\boldsymbol{Z}}_{jl}\left(1+\frac{\lambda}{\rho w_{jl}\|\widehat{\boldsymbol{Z}}_{jl}\|_F} \right),
\end{equation}
and
\begin{equation}
\label{normA}
    \|\boldsymbol{A}_{jl}\|_F = \|\widehat{\boldsymbol{Z}}_{jl}\|_F+\frac{\lambda}{\rho w_{jl}}.
\end{equation}
Finally, by Equation \eqref{normA}, we have
\begin{equation}
\label{claim1cond2}
    \left(\|\boldsymbol{A}_{jl}\|_F-\frac{\lambda}{\rho w_{jl}}\right)_+ > 0.
\end{equation}
Now, we make the following claim.
\begin{enumerate}
    \item $\widehat{\boldsymbol{Z}}_{jl} = 0 \iff \|\boldsymbol{A}_{jl}\|_F \leq \frac{\lambda}{\rho w_{jl}}$.\\
This claim is easily shown by Equations \eqref{claim1cond1} and \eqref{claim1cond2}.
\item When $\|\widehat{\boldsymbol{Z}}_{jl}\|_F \neq 0$, then we have
\begin{equation*}
    \widehat{\boldsymbol{Z}}_{jl} = \boldsymbol{A}_{jl}\left(\frac{\|\boldsymbol{A}_{jl}\|_F-\lambda/(\rho w_{jl})}{\|\boldsymbol{A}_{jl}\|_F} \right).
\end{equation*}
To prove this claim, let us note that from Equation \eqref{claim2cond1} and Equation \eqref{normA} we have
\begin{equation}
\label{zhat}
    \widehat{\boldsymbol{Z}}_{jl} = \boldsymbol{A}_{jl}\left(\frac{\rho w_{jl}\|\widehat{\boldsymbol{Z}}_{jl}\|_F}{\rho w_{jl}\|\widehat{\boldsymbol{Z}}_{jl}\|_F + \lambda} \right),
\end{equation}
and
\begin{equation}
\label{normZ}
        \|\widehat{\boldsymbol{Z}}_{jl}\|_F = \|\boldsymbol{A}_{jl}\|_F-\frac{\lambda}{\rho w_{jl}},
\end{equation}
respectively.
By replacing $ \|\widehat{\boldsymbol{Z}}_{jl}\|_F$ in Equation \eqref{zhat} with the right hand of Equation \eqref{normZ}, the claim is proven.
\end{enumerate}
\noindent It should be noted that, since $\Gamma(\boldsymbol{Z})$ is convex, this algorithm is guaranteed to converge to the global optimum \citep{boyd2011distributed}. Moreover, the positive-definiteness constraint on $\widehat{\boldsymbol{\Theta}}_{0}$ is naturally enforced by the way step 1 is solved.

To solve Equation \eqref{constrfgm}, we start by noticing that it can be reformulated as
\begin{equation*}
    \widehat{\boldsymbol{\Theta}}_{1s} = \argmin_{\boldsymbol{\Theta}}\biggl\{-\log|\boldsymbol{\Theta}| + \Tr\left(\boldsymbol{P}\boldsymbol{S}_n\boldsymbol{P}^T\boldsymbol{\Theta} \right) \biggl\},
\end{equation*}
subject to $\boldsymbol{\Theta} \succ 0$, $\boldsymbol{Z} = \boldsymbol{\Theta}$ and $\boldsymbol{Z}_{jl} = \widehat{\boldsymbol{\Theta}}_{0,jl}$, $(j,l)\notin I(s)$.
Let us note that placing constraints on the elements of $\boldsymbol{Z}$ is equivalent to the constraints in Equation \eqref{constrfgm}.
The scaled augmented Lagrangian \citep{boyd2011distributed} is
\begin{equation}
\label{augmented lag 2}
    L_\rho \left(\{\boldsymbol{\Theta}\},\{\boldsymbol{Z}\},\{\boldsymbol{U}\}\right) = -\log|\boldsymbol{\Theta}| + \Tr\left(\boldsymbol{P}\boldsymbol{S}_n\boldsymbol{P}^T\boldsymbol{\Theta} \right)  + \frac{\rho}{2}||\boldsymbol{\Theta} - \boldsymbol{Z} + \boldsymbol{U}||_F^2,
\end{equation}
where $\rho \geq 0$ is a tuning parameter and $\boldsymbol{U}$ is the dual variable.
The ADMM will solve Equation \eqref{augmented lag 2} by iterating the following steps until convergence. At iteration $i$:
\begin{enumerate}
    \item $\{\boldsymbol{\Theta}_{(i)}\} \leftarrow \argmin_{\{\boldsymbol{\Theta}\}} L_\rho \left(\{\boldsymbol{\Theta}\},\{\boldsymbol{Z}_{(i-1)}\},\{\boldsymbol{U}_{(i-1)}\}\right)$.
    \item $\{\boldsymbol{Z}_{(i)}\} \leftarrow \argmin_{\{\boldsymbol{Z}\}} L_\rho \left(\{\boldsymbol{\Theta}_{(i)}\},\{\boldsymbol{Z}\},\{\boldsymbol{U}_{(i-1)}\}\right)$.
    \item $\{\boldsymbol{U}_{(i)}\} \leftarrow \{\boldsymbol{U}_{(i-1)}\} + \{\boldsymbol{\Theta}_{(i)}\} - \{\boldsymbol{Z}_{(i)}\}$.
\end{enumerate}
The solution $\boldsymbol{\Theta}_{(i)}$ of step 1 can be obtained, as previously described, by replacing $\widehat{\boldsymbol{\Sigma}}_0$ with $\boldsymbol{S}_n$ in Equation \eqref{step1}.

The solution $\boldsymbol{Z}_{(i)}$ of step 2 is obtained as the minimizer of
\begin{align}
\label{step 2 2}
     &\frac{\rho}{2}||\boldsymbol{\Theta}_{(i)} - \boldsymbol{Z} + \boldsymbol{U}_{(i-1)}||_F^2 \\
& \text{s.t.} \quad 
\boldsymbol{Z}_{jl} = \widehat{\boldsymbol{\Theta}}_{0,jl}, \quad (j,l)\notin I(s).
\nonumber
\end{align}
By defining $\boldsymbol{A} = \boldsymbol{\Theta}_{(i)} + \boldsymbol{U}_{(i-1)}$ and changing the sign of the term in the norm, we can rewrite Equation \eqref{step 2 2} as
\begin{align*}
     &\frac{\rho}{2}||\boldsymbol{Z} - \boldsymbol{A}||_F^2 \\
& \text{s.t.} \quad 
\boldsymbol{Z}_{jl} = \widehat{\boldsymbol{\Theta}}_{0,jl}, \quad (j,l)\notin I(s),
\end{align*}
whose solution, for any $1 \leq j,l,\leq p$, is
\begin{equation*}
\widehat{\boldsymbol{Z}}_{jl} = 
    \begin{cases}
        \widehat{\boldsymbol{\Theta}}_{0ij} \quad (j,l)\notin I(s),\\
        \boldsymbol{A}_{jl} \quad (j,l)\in I(s).
    \end{cases}
\end{equation*}

\bibliographystyle{chicago}
{\small
\bibliography{ref}}